\documentclass[fleqn,usenatbib]{mnras}

\usepackage{newtxtext,newtxmath}
\usepackage[T1]{fontenc}
\DeclareRobustCommand{\VAN}[3]{#2}
\let\VANthebibliography\thebibliography
\def\thebibliography{\DeclareRobustCommand{\VAN}[3]{##3}\VANthebibliography}
\usepackage{graphicx}	
\usepackage{amsmath}	
\usepackage{latexsym}
\usepackage{physics}
\usepackage{subfigure}

\title[FDM+gas]{Exploration of simple scenarios involving Fuzzy Dark Matter cores and gas at local scales}

\author[I. Alvarez-R\'ios \& F. S. Guzm\'an]{
Iv\'an \'Alvarez-Rios\thanks{E-mail: ivan.alvarez@umich.mx}
and Francisco S. Guzm\'an\thanks{E-mail: francisco.s.guzman@umich.mx}
\\
Instituto de F\'{\i}sica y Matem\'{a}ticas, Universidad
              Michoacana de San Nicol\'as de Hidalgo. Edificio C-3, Cd.
              Universitaria, 58040 Morelia, Michoac\'{a}n,
              M\'{e}xico.
}

\date{Accepted XXX. Received YYY; in original form ZZZ}

\pubyear{20XX}

\begin{document}
\label{firstpage}
\pagerange{\pageref{firstpage}--\pageref{lastpage}}
\maketitle

\begin{abstract}
We introduce a tool that solves the Schr\"odinger-Euler-Poisson system of equations and allows the study of the interaction between ultralight bosonic dark matter, whose dynamics is described with the Schr\"odinger-Poisson system and luminous matter which, as a first approximation, is modeled with a single component  compressible ideal fluid. The two matter fields  are coupled through the Poisson equation, whose source is the addition of both, dark matter and fluid densities. We describe the numerical methods used to solve the system of equations and present tests for each of the two components, that show the accuracy and convergence properties of the code. As simple possible applications we present some toy scenarios: i) the merger between a core of dark matter with a cloud of gas, ii) the merger of bosonic dark matter plus fluid configurations, and iii) the post merger properties, including the dark-matter offset from gas and the correlation between oscillations of the bosonic core and those of the gas.
\end{abstract}



\section{Introduction}

The model that assumes the dark matter to be an ultralight boson (e.g. \citep{Matos-Urena:2000,Sahni:2000,Hu:2000,Hui:2016}) is currently evolving very rapidly due to its capacity to deal with important problems related to standard cold dark matter (CDM), for example the too-big-to-fail and the cusp-core problems of CDM \citep{Hui:2021tkt,Niemeyer_2020,ElisaFerreira,Lee:2017,Hui:2016,Suarez:2013}. The state of the art of this model includes every day more complex simulations, that detail the evolution of large scale and local dark matter structures, as well in the formation of local structures and their density profiles 
\citep{2009ApJ...697..850W,Schive:2017biq,Schive:2014dra, Schive:2014hza,Mocz:2017wlg,Mocz:2018,mocz19,mocz19b, Schwabe:2016,Shapiro2021}.
These simulations are constructed with codes developed within the community, that use different numerical methods, formulation of equations, platforms and post-processing analysis tools that have boosted research of this promising dark matter candidate, see for example \citep{JiajunZhang,Niemeyer_2020} for reviews of code implementations.
State of the art codes within this dark matter scenario, aside of allowing simulations of structure formation also include baryonic matter dynamics that consider cooling processes suitable for star formation processes  \citep{mocz19,mocz19b,GotingaBarionFDM}.

The essence of simulations and codes is the solution of the Schr\"odinger-Poisson (SP) system of equations, which is the low energy and weak gravitational field version of the Einstein-Klein-Gordon equations, expected to rule the dynamics of a spin-less boson at early universe scales, and is also identified with the evolution system of a Bose Einstein Condensate (BEC) in the Gross-Pitaevskii regime \citep{grosspitaevskii}, subject to a gravitationally self-generated trap. The most studied case of the SP system is the free field case, in which no self-interaction between bosons is considered, the so called Fuzzy Dark Matter (FDM) regime. Even though in FDM no self-interaction is considered, BEC dark matter and FDM are used indistinctly, and so we do in this paper.

At local scales some problems on the dynamics of FDM cores have been studied, for example the merger of cores \citep{Schwabe:2016,Maleki_2020,GuzmanAlvarezGonzalez2021}, the structure of dark matter halos \citep{Du_2016,YavetzLiHui2022}, the  interaction of dark matter halos and orbiting bodies \citep{Ferreira}, the dynamics of rotating blobs of dark matter \citep{RotatingDrops}, core-halo mass relations from binary mergers \citep{Du2017}, the disruption of sub-halos \citep{PhysRevD.97.063507}, the inclusion of supermassive black holes in the center of cores \citep{Moczfdmbh}, the heating effects of dark matter on galactic disks \citep{Moczheating}, among others that can trigger  a number of interesting scenarios to be compared with observations.

The model has a number of areas of opportunity and constraints. While the special signature of this dark matter candidate is its ultralight mass, which helps preventing the formation of excessive dwarf structures due to its large de Broglie wavelength, as well the formation of cuspy cores, its consistency with observations on local scenarios is under permanent search of constraints. Among the most studied effects is the restless nature of this type of matter, that forms cored but permanently oscillating structures, surrounded by a background of matter interfering at all times with luminous matter. Observational predictions, as well as constraints on the model expected to arise, go together with possible drawbacks  mostly related to the possible effects that the restless behavior of the boson density can have on luminous structures, like stelar clusters near galactic cores, that could be overheated or even disrupted \citep{Marsh_2019,GotingaBarionFDM}. On the contrary, that tidal forces may lead to heating suppression of star clusters \citep{Schive_2020,Li_2021}. On smaller scales, the effects on compact objects, like consistency of ultralight bosonic matter with spin observations in the supermassive black holes of M87 \citep{EHT1} introduce restrictions to this dark matter particle  \citep{Davoudiasl_2019}. Dynamical friction is a property that comes with the model \citep{Hui:2016} and should affect the motion of satellite galaxies, for example Fornax \citep{Lancaster_2020}. Also, friction is related to diffusion \citep{Bar_Or_2019}, effects that can eventually contradict observations or restrict the model.

Motivated by this rich variety of interesting scenarios worth investigating, that couple the dynamics of dark matter and luminous matter, we present a code that solves the Schr\"odinger-Euler-Poisson (SEP) system at local scales, that couples the SP to Euler equations. We have developed this code based on the assumption that Euler equations can model the dynamics of luminous matter whereas the SP rules the dynamics of dark matter. The code results from merging CAFE, a code that solves Euler and Magnetohydrodynamics equations \citep{NewtonianCAFE}, with our code that solves the 3D SP system \citep{GuzmanAlvarezGonzalez2021}, through Poisson equation, which is sourced by dark matter and fluid densities. 


We consider simple scenarios that test the accuracy and convergence properties of the numerical implementation, as well as scenarios that show the dynamics of the interaction between FDM cores and clouds of gas. In particular we introduce a proof of principle of how dark matter-baryon configurations could form, by assuming the existence of dark matter potential wells due to the attractor formation properties of equilibrium solutions of the SP system, which can interact with clouds of gas traveling around. 
A second explored scenario involves the construction of equilibrium configurations of both dark matter and gas, that we use to emulate mergers of dark matter-baryon configurations. We practice this test in head-on and orbital motion and study the post-merger properties of the configurations. These include the stationarity of the resulting final configuration made of FDM and gas, together with the relation between the oscillations of dark matter with those of the gas.
With the illustrative examples in this paper, we expect the code to be useful in the study of the bosonic dark matter phenomenology.

The paper is organized as follows. In Section \ref{sec:model} we detail the numerical methods used to solve the equations and  tests of the code, whereas in Section \ref{sec:applications} we present some applications of the code. Finally in Section \ref{sec:conclusions} we present some final comments.

\section{The code}
\label{sec:model}

\subsection{The system of equations}

The BEC+fluid system is considered to obey the SPE equations, coupled through gravity. This means that both matter components generate their own gravitational field and  interact through gravity as expected for the interaction between dark and luminous matter. The SPE system is written as

\begin{eqnarray}
i\hbar \frac{\partial \tilde{\Psi}}{\partial \tilde{t}} &=& -\frac{\hbar^2}{2m_B}\tilde{\nabla}^2\tilde{\Psi} + m_{B}\tilde{V}\tilde{\Psi}\label{eq:GPP1}\\
\frac{\partial \tilde{\rho}}{\partial \tilde{t}} &+& \tilde{\div}\left(\tilde{\rho}\tilde{\vec{v}}\right)=0,\label{eq:EGPPrho2}\\
\frac{\partial \left(\tilde{\rho} \tilde{\vec{v}}\right)}{\partial \tilde{t}} &+&\tilde{\div}\left(\tilde{\rho}\tilde{\vec{v}}\otimes\tilde{\vec{v}}+\tilde{p} \vb{I}\right)=-\tilde{\rho}\tilde{\grad} \tilde{V},\label{eq:EGPPj2}\\
\frac{\partial \tilde{E}}{\partial \tilde{t}} &+&\tilde{\div}\left[\tilde{\vec{v}}\left(\tilde{E}+\tilde{p}\right)\right]=-\tilde{\rho}\tilde{\vec{v}}\cdot\tilde{\grad} \tilde{V},\label{eq:EGPPE2}\\
\tilde{\nabla}^2\tilde{V} &=& 4\pi G \left(\tilde{\rho} + m_B|\tilde{\Psi}|^2\right),\label{eq:GPP2}
\end{eqnarray}

\noindent where each fluid volume element is characterized by its mass density $ \tilde{\rho} $, velocity $ \tilde{\vec {v}} $, internal energy $ \tilde{e} $, pressure $ \tilde{p}$, and total energy $\tilde{E}=\tilde{\rho}(\tilde{e}+\frac{1}{2}|\tilde{\vec{v}}|^2)$. The bosonic structure is described via the order parameter $\tilde{\Psi}$, and the two components are coupled to the gravitational potential $\tilde{V}$. Rescaling  the variables with the rules $\Psi=\frac{\sqrt{G m_B^3} R_0^2}{\hbar}\tilde{\Psi}$, $\rho = \frac{G m_{B}^2 R_0^4}{\hbar^2} \tilde{\rho}$, $\vec{v} = \frac{m_B R_0}{\hbar} \tilde{\vec{v}}$, $e = \left(\frac{m_B R_0}{\hbar}\right)^2 \tilde{e}$, $p = \frac{G m_B^4 R_0^6}{\hbar^4}\tilde{p}$, $E = \frac{G m_B^4 R_0^6}{\hbar^4} \tilde{E}$, $V=\left(\frac{m_{B}R_0}{\hbar}\right)^2\tilde{V}$,  and coordinates $x=\frac{\tilde{x}}{R_0}$, $y=\frac{\tilde{y}}{R_0}$, $z=\frac{\tilde{z}}{R_0}$, $t=\frac{\hbar}{m_B R_0^2}\tilde{t}$, where $R_0 = \frac{\hbar^2}{G m_{B}^2 M_0}$ is the relation between the arbitrary mass scale $M_0$ and the length scale $R_0$. In terms of the new variables, the constants $G,\hbar,m_B$ are absorbed and the SPE system of equations is transformed into:

\begin{eqnarray}
\frac{\partial \rho}{\partial t} &+& \div\left(\rho\vec{v}\right)=0,\label{eq:EGPPrho}\\
\frac{\partial \left(\rho \vec{v}\right)}{\partial t} &+&\div\left(\rho\vec{v}\otimes\vec{v}+p \vb{I}\right)=-\rho\grad V,\label{eq:EGPPj}\\
\frac{\partial E}{\partial t} &+&\div\left[\vec{v}\left(E+p\right)\right]=-\rho\vec{v}\cdot\grad V,\label{eq:EGPPE}\\
i\frac{\partial \Psi}{\partial t} &=& -\frac{1}{2}\nabla^2 \Psi + V\Psi,\label{eq:EGPPSchro}\\  
\nabla^2 V &=& 4\pi(  |\Psi|^2 + \rho),\label{eq:EGPPPoisson}
\end{eqnarray}

\noindent which is the system to be solved numerically. This is a set of seven equations for eight variables $\rho, \vec{v}, e, p,\Psi,V$, that we close using an equation of state. In one case we use the ideal gas equation of state $p = \left(\gamma-1\right)\rho e$, with $\gamma$ the adiabatic index of the gas and in a second case we set $p=0$ in order to consider dust.

This constrained evolution problem is solved provided consistent initial conditions on the fluid variables and the order parameter of the BEC as follows

\begin{eqnarray}
\rho(\vec{x},t=0) &=& \rho_0(\vec{x}),\nonumber\\
\vec{v}(\vec{x},t=0)&=&\vec{v}_0(\vec{x}),\nonumber\\
p(\vec{x},t=0)&=&p_0(\vec{x}),\nonumber\\
\Psi(\vec{x},t=0) &=& \Psi_0(\vec{x}).\nonumber
\end{eqnarray}

\noindent From the pressure, either the ideal gas or dust, one calculates the internal energy at initial time $e_0(\vec{x})=\frac{p_0}{\rho_0(\gamma-1)}$ and finally the total energy $E_0(\vec{x})$, which define the initial conditions for the variables of the system. Once these variables are defined one can integrate Poisson equation to calculate $V$ at initial time, which completes the initial conditions.

\subsection{Numerical methods}

The system of equations (\ref{eq:EGPPrho})-(\ref{eq:EGPPPoisson}) has five hyperbolic-type evolution equations for the fluid, a non-hyperbolic complex equation for $\Psi$ and a constraint elliptic equation for $V$ that has to be satisfied during the evolution.

We solve these equations as an initial value problem on the domain $D\times [0,t_f]:=[x_{min},x_{max}]\times[y_{min},y_{max}]\times[z_{min},z_{max}]\times[0,t_f]$ described with Cartesian coordinates. 
We define the discrete version of the domain as $D_d = \{ (x_i,y_j,z_k)\in D | x_i=x_{min}+i\Delta x, y_j=y_{min}+j\Delta y, z_k=z_{min}+k\Delta z \}$, where $i=0,...,N_x$, $j=0,...,N_y$, $k=0,...,N_z$ are the number of cells along each direction and $\Delta x=(x_{max}-x_{min})/N_x$, $\Delta y=(y_{max}-y_{min})/N_y$, $\Delta z=(z_{max}-z_{min})/N_z$ are the numerical resolutions. 
In the examples below we will always use the same numerical resolution along each direction $\Delta x=\Delta y=\Delta z$. 

Since we expect the code to be applied in the study of local phenomena, the use of Fixed Mesh Refinement (FMR) is helpful to concentrate the accuracy in the region near the center of mass of the system. We use concentric refined boxes defined at the center of the domain, that is, the domain with refinement level $l$ is $D^l_d = [\frac{x_{min}}{2^{l-1}},\frac{x_{max}}{2^{l-1}}]\times[\frac{y_{min}}{2^{l-1}},\frac{y_{max}}{2^{l-1}}]\times[\frac{z_{min}}{2^{l-1}},\frac{z_{max}}{2^{l-1}}]$ with spatial resolution $\Delta x^l = \frac{\Delta x}{2^{l-1}}$. For the evolution of the variables across time levels we use the Berger-Oliger method \citep{berger1984amr}.

The evolution of $\Psi$ and the fluid variables uses the Method of Lines (see e.g. \citep{gustafsson1995time}) for the system (\ref{eq:EGPPrho})-(\ref{eq:EGPPSchro}) with, a third order Runge-Kutta integrator (RK3). When solving Schr\"odinger equation alone, the MoL uses time resolution $\Delta t = C \Delta x^2$, whereas when solving Euler equations we use $\Delta t=C\Delta x$, where $C$ is the CFL factor. In the SEP system we use the more restrictive one $\Delta t = C \Delta x^2$.

The subsystem of Euler equations (\ref{eq:EGPPrho})-(\ref{eq:EGPPE}) is solved using a High Resolution Shock Capturing method based on a Finite Volume discretization of the domain, together with Marquina and HLLE \citep{hlle1,hlle2,Toro2009} numerical fluxes. Equations (\ref{eq:EGPPrho})-(\ref{eq:EGPPE}) are already written in flux balance form, so that the vector of conserved variables is ${\bf u}=[\rho,\rho \vec{v},E]^T$, whereas the fluxes are the arguments of the divergence operators and the sources are the right hand sides of these equations. For the reconstruction of variables in the examples of this paper we use the minmod reconstructor, although other linear and higher order are also implemented. In order to avoid divergences we use an atmosphere value to density $\rho_{atm}$ that we set to $10^{-15}$ in all examples in this paper, but still a parameter subject to changes in other scenarios. All these methods together are first order accurate when  discontinuities are formed, whereas second order when all variables are smooth. In general the accuracy of the method lies within first and second order.  Finally, for the scenarios in this paper we use outflow boundary conditions in all conserved variables, however this condition can be relaxed for other cases.

In order to synchronize the evolution with hydrodynamics, Schr\"odinger equation is solved also using the Method of Lines with the same RK3, and a Finite Differences approximation of the right hand side of (\ref{eq:EGPPSchro}). The accuracy of spatial operators can be second, fourth or sixth, however since the methods used for the evolution of the fluid are at most second order, and the use of more accurate stencils does not improve the accuracy of the whole system, we use second order accuracy as default. Considering the system to be isolated, we impose a sponge near the boundary of the domain by using an imaginary potential into Schr\"odinger equation that absorbs the wave function when approaching the boundary as in \citep{GuzmanUrena2004,Grupo2014}. This is an important distinction with codes solving the SP system in structure formation scenarios, where periodic boundary conditions are used that retain the matter within the domain.

Finally, Poisson equation (\ref{eq:EGPPPoisson}) is solved using a multigrid scheme with a two step V-cycle consisting of  Successive Overrelaxation substeps \citep{multigrid}. Boundary conditions assume the BEC and the fluid are isolated, for which we set by default a monopolar boundary condition $V_{boundary}=-M_T/r_{boundary}$, where $M_T$ is the total mass of the BEC and the fluid together, and $r_{boundary}$ is the distance from the center to each point in the boundary of the numerical domain. 

\subsubsection{Diagnostics}

A basic set of quantities describing the dynamics of the BEC and the fluid, which are worth monitoring and tracked by the code are for the BEC, its mass $M_{BEC}$, the integrated kinetic energy calculated as the expectation value of the kinetic energy operator in Schr\"odinger equation $K_{BEC}=\langle -\frac{1}{2}\nabla^2 \rangle$,  gravitational energy $W_{BEC}=\frac{1}{2}\langle V\rangle$ and the angular momentum along the $z-$axis useful to diagnose mergers $L_{z,BEC}=\langle L_z \rangle$:

\begin{eqnarray}
M_{BEC} &=& \int|\Psi|^2 d^3 x,\label{eq:BECM}\\
K_{BEC} &=& -\dfrac{1}{2}\int \Psi^* \nabla^2 \Psi d^3 x, \label{eq:BECK}\\
W_{BEC} &=& \dfrac{1}{2}\int |\Psi|^2 V d^3 x,\label{eq:BECW}\\
L_{z, BEC} &=& -i \int \Psi^*\left(x\dfrac{\partial \Psi} {\partial y} - y \dfrac{\partial \Psi}{\partial x}\right) d^3 x.\label{eq:BECLz}
\end{eqnarray}

\noindent The equivalent global quantities for the gas are calculated as follows

\begin{eqnarray}
M_{gas} &=& \int \rho d^3 x, \label{eq:gasmass}\\
K_{gas} &=& \dfrac{1}{2}\int \rho |\vec{v}|^2 d^3 x,\label{eq:gasKenergy}\\
W_{gas} &=& \dfrac{1}{2}\int \rho V d^3 x,\label{eq: gasWenergy}\\
L_{z, gas} &=& \int \rho \left(x v_{y} - y v_{x}\right) d^3 x. \label{eq:gasLz}
\end{eqnarray}

\noindent Important additional quantities for the BEC and the fluid are their total energy $E_{BEC}=K_{BEC}+W_{BEC}$, $E_{gas}=K_{gas}+W_{gas}$ and the monitor function for virialization $2K_{BEC}+W_{BEC}$, $2K_{gas}+W_{gas}$, which is zero for a virialized system.

Besides the diagnostics of physical quantities, in some of the tests below we estimate the $L_2$ and infinity norms of grid functions, specifically errors or difference among numerical solutions. These norms for an arbitrary grid function $f$ defined in a given numerical domain $D_d$ are:

\begin{eqnarray}
L_2(f)=||f||_2 &=& \int_{D_d} f ~d^3 x,\label{eq:norm2}\\
L_{\infty} =||f||_{\infty} &=& \max (f(x_i,y_j,z_k)),\label{eq:norminfty}
\end{eqnarray}

\noindent for all $(x_i,y_j,z_k)\in D_d$.

\subsection{Information about the code}

The code is programmed in fortran 90 and no external libraries are used for the implementation of the numerical methods. We use the MPI library for parallelization with standard domain decomposition. Output is produced in ascii for scalars and one dimensional output, whereas the external library HDF5 is used for the output of two and three dimensional information. The simulations in the following sections, being local systems less demanding than structure formation scenarios of FDM where massive parallel processing and high power computing is needed, can be evolved using shared memory computers with a modest number of cores.

\subsection{Tests}

We consider three scenarios, a first one focused on the evolution of the FDM alone, a second one oriented to monitor accuracy in the  evolution of the fluid and a third one that has both components.

\subsubsection{BEC}

This is the standard test for the SP system, and consists in the correct evolution of a ground state equilibrium solution of the system, with the standard numerical parameters of mass $M\sim 2.06$ and radius $r_{95}\sim 3.93$ \citep{GuzmanUrena2004}. We interpolate the solution constructed using spherical coordinates according to \citep{GuzmanUrena2004} into our 3D cartesian grid in the domain $D=[-20,20]\times[-20,20]\times[-20,20]$. Since the solution is stationary, the density of the BEC should remain time independent in the continuum limit whereas the wave function must evolve with the appropriate frequency. Therefore tracking a scalar of the density is helpful to verify whether the evolution is correct, for which we use the central density of the configuration as function of time.

\begin{figure}
\centering
\includegraphics[width=8cm]{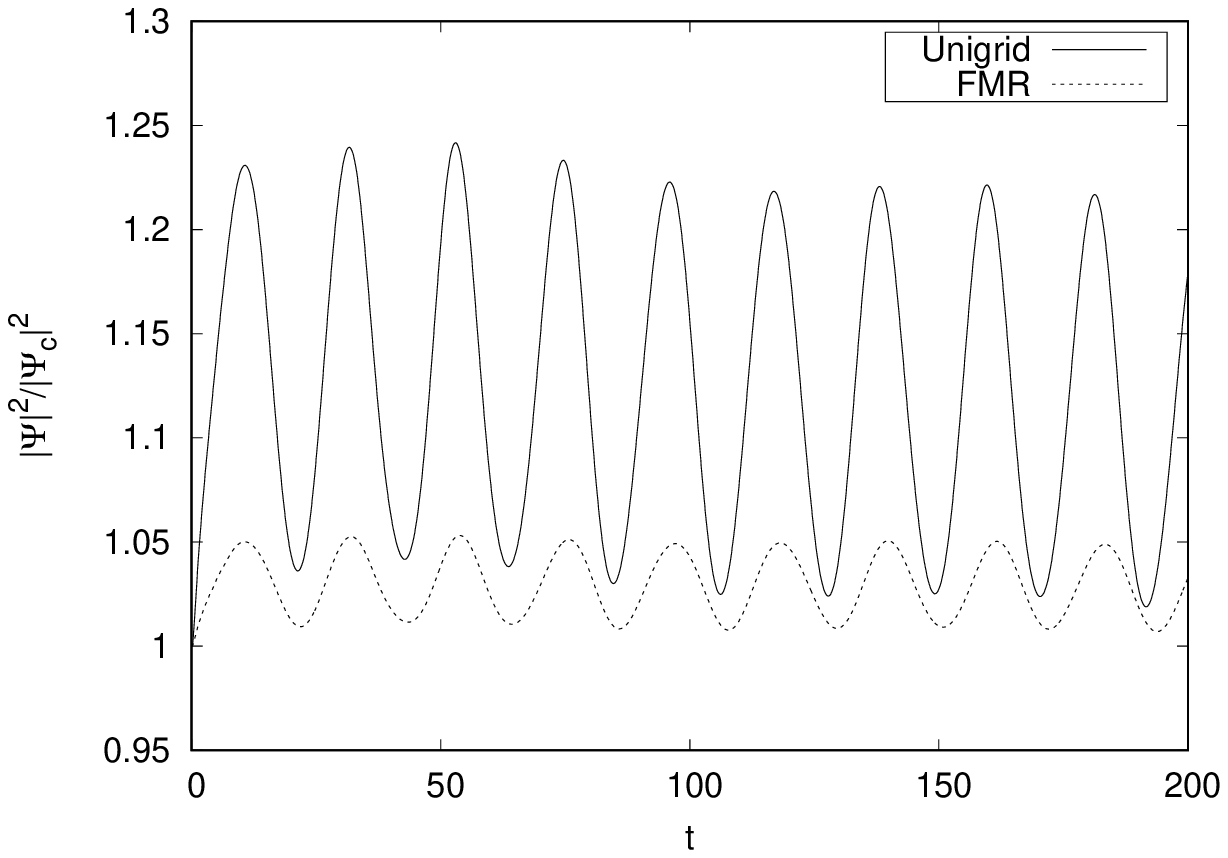}
\includegraphics[width=8cm]{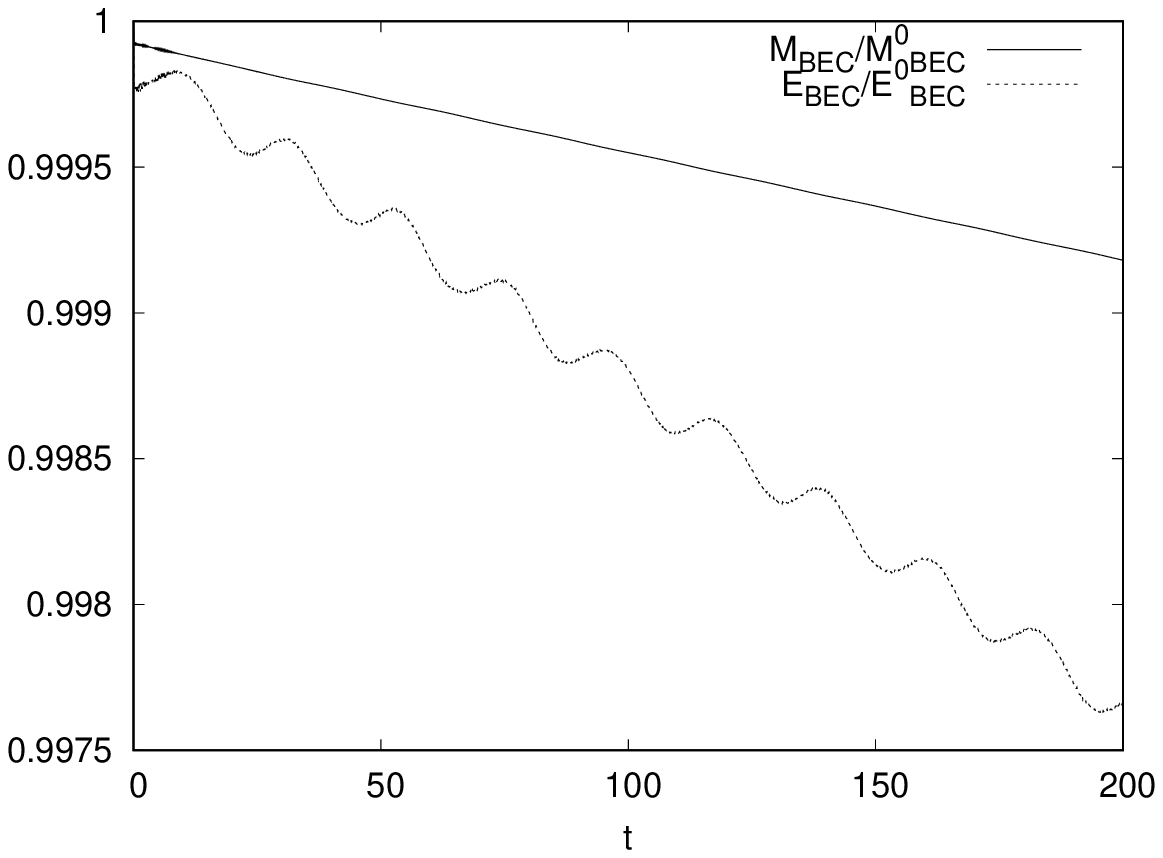}
\caption{\label{fig:BECeq} (Top) Central density $|\Psi|^2$ as function of time for the equilibrium configuration using unigrid and FMR, normalized to its initial value. In the continuum, this quantity should remain constant and equal to one. The solution using the unigrid mode with coarse resolution shows oscillations four times bigger than the oscillations using the FMR with double resolution covering the center of the domain, which shows second order convergence in the infinity norm of the density error.  (Bottom) Mass $M_{BEC}$ and total energy $E_{BEC}=K_{BEC}+W_{BEC}$ as in expressions (\ref{eq:BECM},\ref{eq:BECK},\ref{eq:BECW}), calculated with the FMR mode. Within the time window the loss in mass is of order 0.05\%,  which expresses the deviation from unitarity, whereas the lost portion of energy is of order 0.2\%. }
\end{figure}

The result is shown in Figure \ref{fig:BECeq}. In one case we use the code in unigrid mode with resolution 
$\Delta x=\Delta y=\Delta z=0.5$. In a second case we use the FMR mode and discretize the subdomain $[-10,10]\times[-10,10]\times[-10,10]$ with resolution $\Delta x=\Delta y=\Delta z=0.25$. In Figure \ref{fig:BECeq} the amplitude of the oscillations are nearly four times bigger using unigrid than the amplitud with the FMR covering the central subdomain with double resolution, which indicates second order convergence of the infinity norm to the exact value in the continuum \citep{gustafsson1995time}.

Also in Figure \ref{fig:BECeq} we test the conservation of mass and energy using the run with the high resolution. The conservation of mass is an indication of the unitarity of the evolution of the wave function, and this plot shows the loss rate of unitarity due to numerical dissipation inherent to numerical evolution. We show the loss rate within the time window used for the convergence test.

\subsubsection{Fluid}

The test for the fluid part is the evolution of a Newtonian Tolman-Oppenheimer-Volkov (TOV) star model which couples Euler equations of the fluid to its own gravitational potential $V$. For the construction of this star model, we follow \citep{ShapTeu} and solve the Lane-Emden equation \\

\begin{equation}
\dfrac{1}{\xi^2}\dfrac{d}{d\xi}\left(\xi^2\dfrac{d\theta}{d\xi}\right) = - \theta^n, 
\label{eq: Lane-Emden}
\end{equation}

\noindent with the boundary conditions $\theta(0) = 1$, $\theta(\xi_1) = 0$ and $\frac{d\theta}{d\xi}(0)=0$, where $\xi_1$ is interpreted as the star radius. The function $\theta$ is related to the density  of the fluid through the expression $\rho = \rho_c \theta^n$ where $\rho_c$ is the central density of the TOV star. The pressure is calculated with the polytropic equation of state $p = K \rho^\gamma$ where $K$ is the polytropic constant, $n$ the polytropic index and $\gamma := 1 + 1/n$ is the adiabatic index.  The radial coordinate is given by $r := \alpha \xi$ where $\alpha^2 = K(n+1)\rho_c^{\frac{1-n}{n}} / (4\pi)$. \\

The total mass $M$ and  radius $R=\alpha \xi_1$ of the star are related by the expression 

\begin{equation}
R^{\frac{3-n}{n}}M^{\frac{n-1}{n}} = \dfrac{K}{G N_n}, 
\label{eq: RM TOV}
\end{equation}

\noindent where

\begin{equation}
N_n := \dfrac{(4\pi)^{1/n}}{n+1}\left(\left[-\xi^2\dfrac{d\theta}{d\xi}\right]_{\xi=\xi_1}\right)^{\frac{1-n}{n}} \xi_1^{\frac{n-3}{n}},
\label{eq: Nn}
\end{equation}

\noindent and the central density can be written as

\begin{equation}
\rho_c = \dfrac{M}{4\pi R^3} \dfrac{\xi_1^3}{\left[-\xi^2\dfrac{d\theta}{d\xi}\right]_{\xi=\xi_1}}.
\label{eq: rhoc}
\end{equation}

\begin{figure}
\centering
\includegraphics[width=8cm]{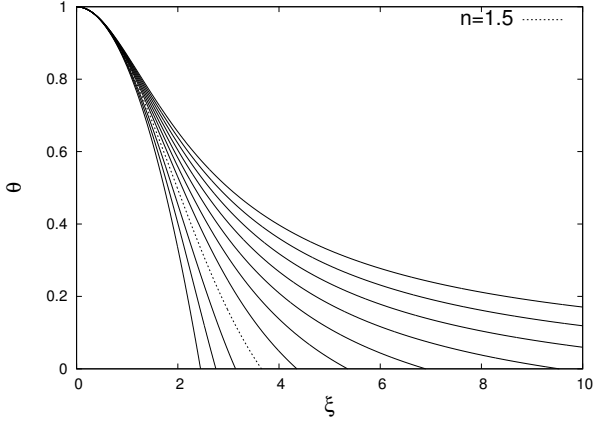}
\caption{Some solutions of the Lane-Emden equation for various values. The curves from left to right correspond to 
 $n=0.5\times i$, for $i=0,...,10$. A particular example for $n=1.5$ is shown with the dotted line, a case that will be used later on.}
\label{fig: Lane-Emden}
\end{figure}

\noindent In Figure \ref{fig: Lane-Emden} we show various solutions of the Lane-Emden equation corresponding to various values of $n$ or equivalently $\gamma$. In the particular case of a monoatomic ideal gas $\gamma=5/3$, the solution is drawn with dotted line. In order to construct the TOV solution one has to provide the mass $M$ and  radius $R$, or equivalently the central density $\rho_c$ and the polytropic constant $K$. 
For the test, we set the mass of the TOV star to $M=0.206$ and radius $R=15$, which determines uniquely the TOV solution to have a central density $\rho_c \approx 1.428 M / R^3$ and $ K \approx 0.4242 G R M^{1/3}$.

We then interpolate the density $\rho$ of the star into the 3D domain in Cartesian coordinates. Prior to start the evolution we populate the domain with all other hydrodynamical variables $p$ in terms of the polytropic equation of state, the internal energy in terms of the ideal equation of state $e=\frac{p}{\rho(\gamma-1)}$, set the stationarity condition ${v_0(\vec{x})}=0$, calculate the total energy $E$ and solve Poisson equation in order to integrate the initial gravitational potential $V$ in the 3D domain.

\begin{figure}
\centering
\includegraphics[width=8cm]{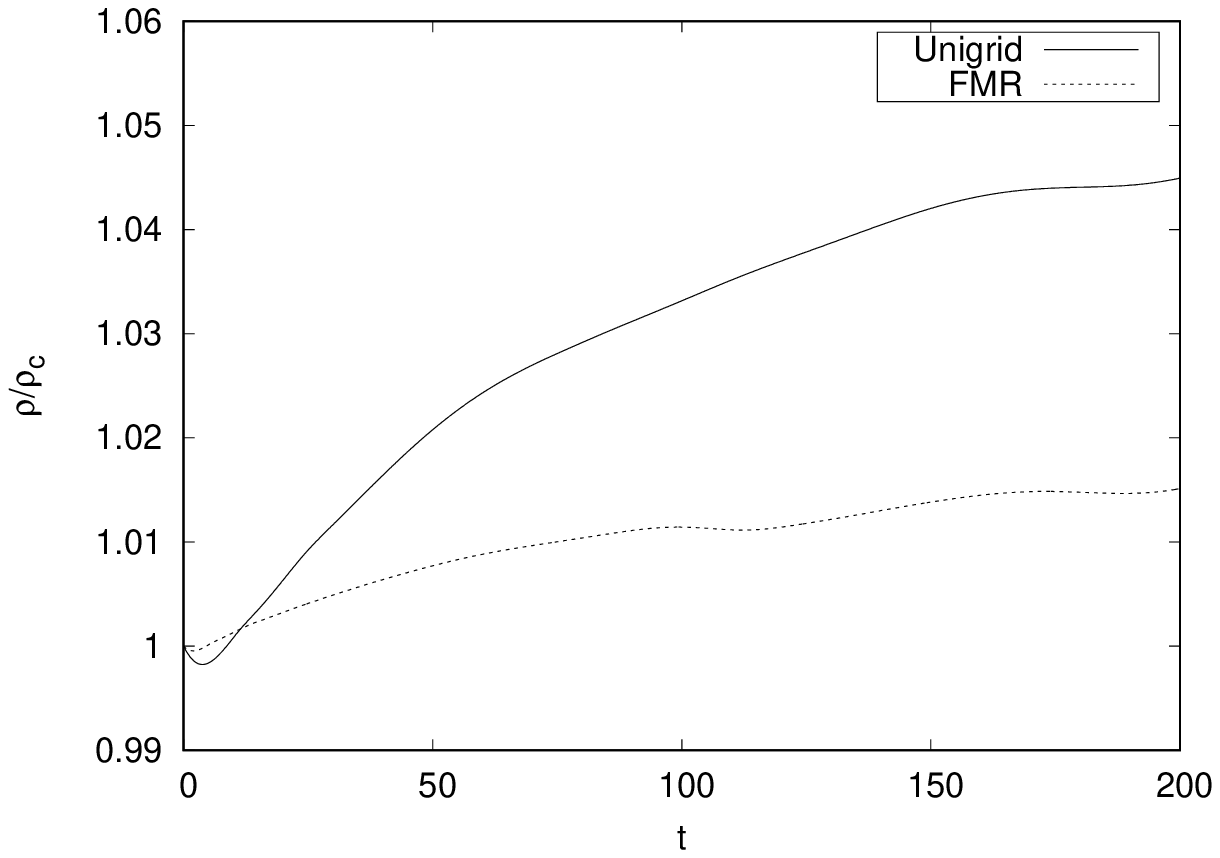}
\includegraphics[width=8cm]{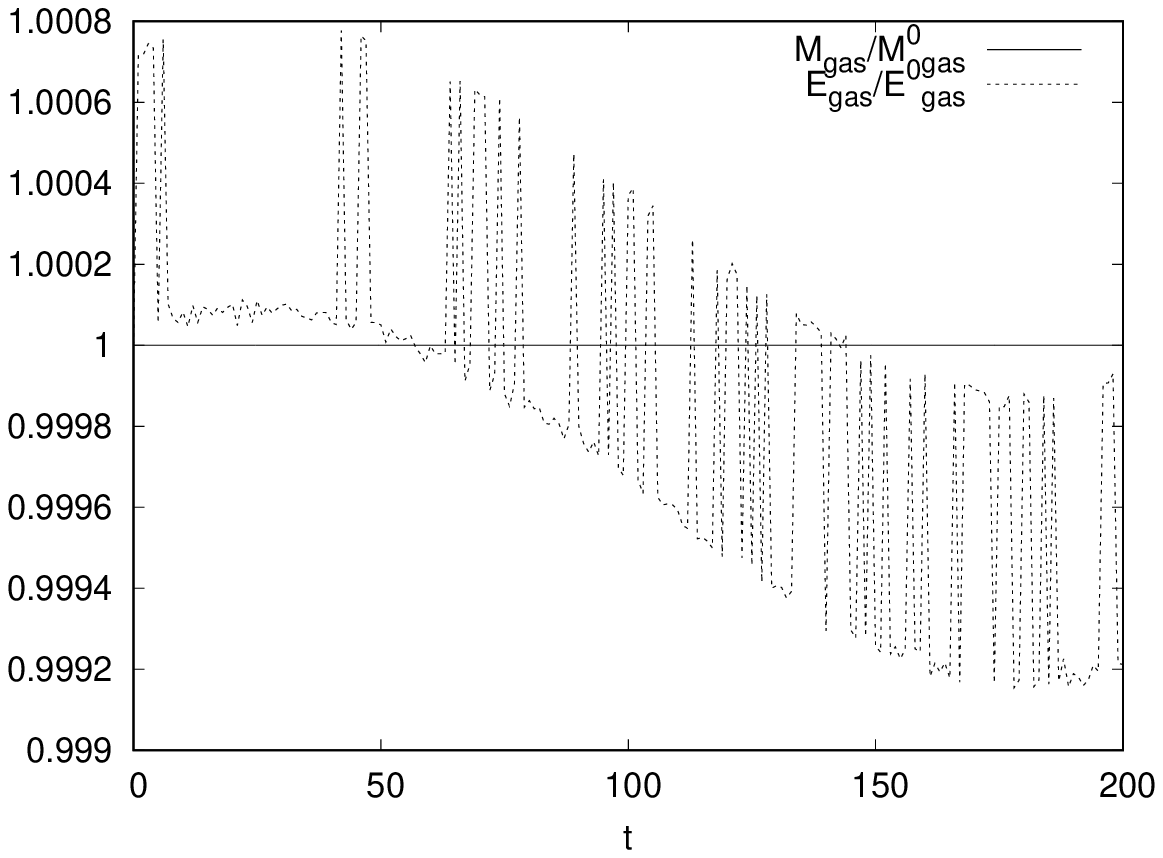}
\caption{\label{fig:TOV} (Top) Central density $\rho(0,0,0)$ as function of time for the TOV equilibrium configuration using unigrid and FMR modes. Departure from 1 is the infinity norm of the error in the numerical value of $\rho$ and we notice that the convergence factor remains between 2 and 4, which indicates a convergence of order between 1 and 2, which is the order of accuracy of our method to solve Euler equations. (Bottom) We show the mass of the fluid $M_{gas}$ and the total energy of the configuration $E_{gas}=K_{gas}+W_{gas}$ as function of time, according to formulas (\ref{eq:gasmass},\ref{eq:gasKenergy},\ref{eq: gasWenergy}). The loss of mass is less than 0.001\% and energy changes by 0.08\%, which indicates how conservative the methods for hydrodynamics are.}
\end{figure}

The numerical domain is exactly the same as that used for the BEC equilibrium configuration, namely $D=[-20,20]\times[-20,20]\times[-20,20]$ with the star at the coordinate center. We simulate the evolution using  unigrid mode with resolution $\Delta x=\Delta y=\Delta z=0.5$ and  FMR mode with the subdomain  $[-10,10]^3$ covered with resolution $\Delta x=\Delta y=\Delta z=0.25$ in order to test the convergence of the central value of the density to its exact value in the continuum. The error of the central density as function of time appears in Figure \ref{fig:TOV} for the unigrid and FMR modes. Unlike for the BEC configuration, in this case the convergence factor lies between 2 and 4, which is consistent with the accuracy of the method to solve Euler equations, which is second order for smooth state variables and first order when discontinuities form. In our set up, the surface of the star at initial time defines a non-smooth density and pressure surface, whose errors propagate across the domain.

Also in Figure \ref{fig:TOV} we show the loss of mass and energy of the gas, using the run with the high resolution. The loss is due to the numerical methods and its accumulation should be considered during evolutions.

\subsubsection{BEC+fluid}

As an example of interaction between BEC and fluid, we study the head-on collision of a TOV star with $\gamma=5/3$, mass $M =2.06$ and radius $R=5$, with the ground state  equilibrium configuration of the first test, with mass $M=2.06$ and radius $r_{95}=3.93$. The initial centers of these objects are located at $(\pm 10,0,0)$ and the head-on velocity is $\pm 1$ along the $x-$axis. 

\begin{figure}
\includegraphics[width=8cm]{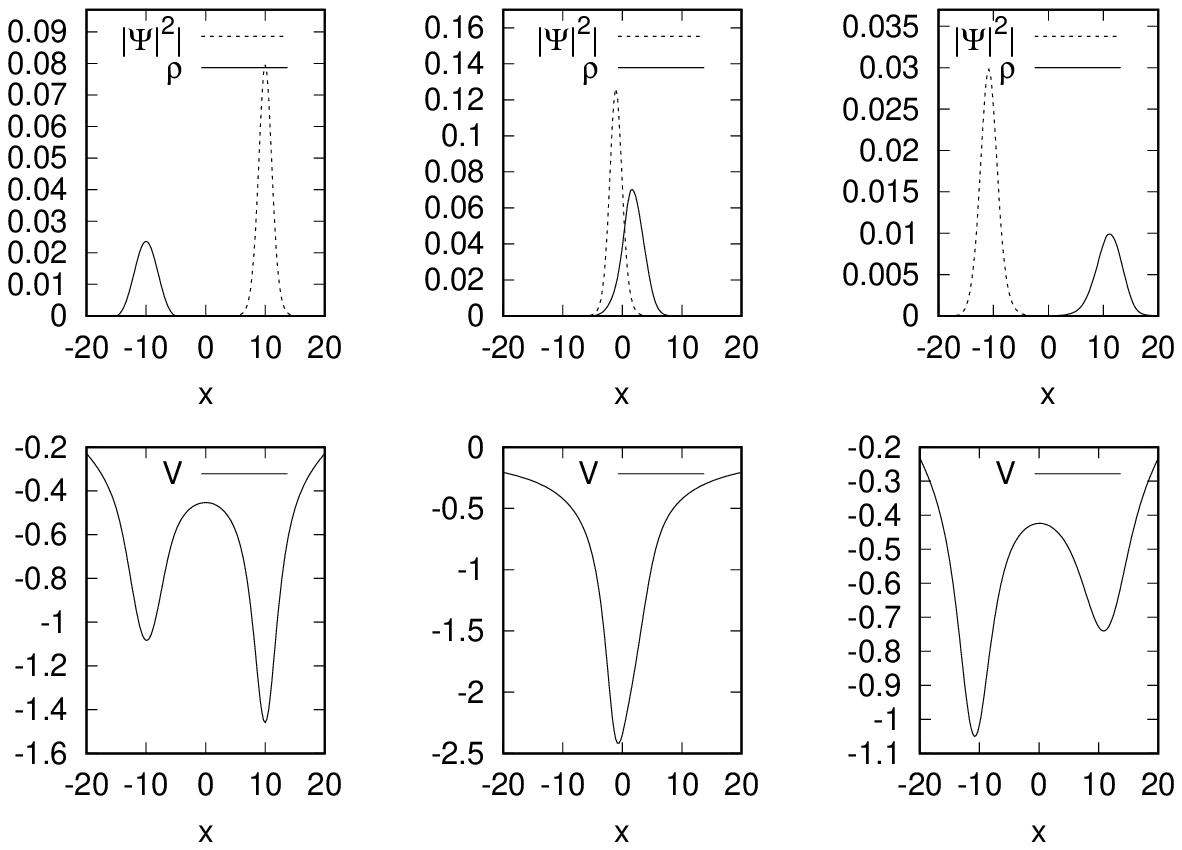}
\includegraphics[width=8cm]{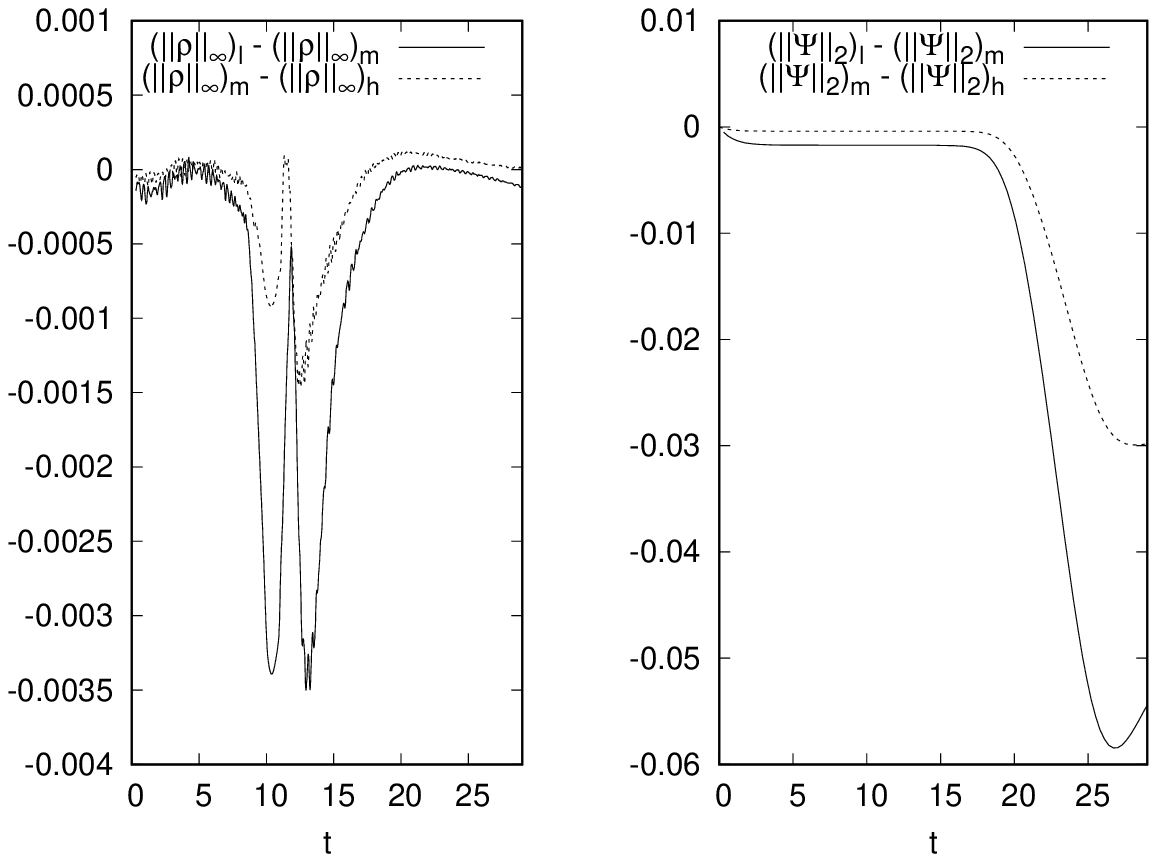}
\caption{\label{fig:test3} (First and second rows) Snapshots of $\rho$, $|\Psi|^2$  and gravitational potential $V$ at times $t=0,10,20$ along the collision axis. (Bottom row) We show $y_l-y_m$ and $y_m-y_h$ where $y$ is $||\rho ||_{\infty}$for the fluid and $||\Psi ||_2$ for the BEC. The factor between the curves is between  $2.07$ and $2.45$, which shows self-convergence according to formula (\ref{eq:scfactor}), in two different norms for the densities of each component.}
\end{figure}

The total energy of the system is positive and therefore the configuration is unbounded. The evolution can be summarized as follows. Density functions $|\Psi|^2$ and $\rho$ interfere, then trespass each other as shown in the snapshots of Figure \ref{fig:test3} and leave the numerical domain.

Unlike previous tests, the solution of this problem is not stationary, there is no prescription on how densities should behave and consequently there is no property of the system that can be used as a reference solution in order to practice convergence tests. In such situation one has to practice self-convergence tests, which consist in the comparison of the numerical solution against itself using different resolutions. For the test we calculate the numerical solution in the domain $[-20,20]\times[-10,10]\times[-10,10]$, that we discretize using high $\Delta x_h = 40 / 144$, medium $\Delta x_m = 40 / 128$ and low $\Delta x_l = 40 / 104$ resolutions. We estimate self-convergence as follows. Let $r_1=128/104$ be the ratio between low and medium resolution and $r_2=144/128$ the ratio between medium and high resolution. Assuming that we calculate solutions $y_l,y_m,y_h$ using respectively low, medium and high resolutions and that the exact solution is $y_e$. Then each of the numerical solutions can be written as

\begin{eqnarray}
y_l&=&y_e + {\cal E} \Delta x_l^n,\nonumber\\
y_m&=&y_e + {\cal E} \left( \frac{\Delta x_l}{r_1} \right)^n,\nonumber\\
y_h&=&y_e + {\cal E} \left( \frac{\Delta x_l}{r_1 r_2} \right)^n,\nonumber
\end{eqnarray}

\noindent where $n$ is the accuracy of the numerical method used to calculate the solutions and ${\cal E}$ is the error of the solution. By subtracting the second expression from the first and the third from the second we obtain

\begin{eqnarray}
y_l - y_m &=& {\cal E} \Delta x_l^n\left(1-\frac{1}{r_1^n}\right),\nonumber\\
y_m-y_h &=& {\cal E} \left( \frac{\Delta x_l}{r_1} \right)^n\left(1-\frac{1}{r_2^n}\right),~~~~~\Rightarrow\nonumber\\
SCF&=&\frac{y_l - y_m}{y_m-y_h}=r_2^n\frac{\left(r_1^n-1\right)}{\left(r_2^2-1\right)},\label{eq:scfactor}
\end{eqnarray}

\noindent which indicates that for $r_1=1.231$ and $r_2=1.125$ with a method first and second order accurate $n=1,~2$, this self-convergence factor (SCF) is respectively $SCF\simeq 2.07$ and $2.45$.

We illustrate the self-convergence in the bottom of Figure \ref{fig:test3}, where we show $y_l-y_m$ and $y_m-y_h$, for $y$ the infinity norm of the fluid density $||\rho||_{\infty}$ as function of time. For the wave function we show $y$ as the $L_2$ norm of $\Psi$, that is, $||\Psi ||_{2}$ as function of time. We use the $L_2$ norm for the squared wave function because its distribution is smooth and can be arbitrarily small during the evolution, then a norm resulting from an integral in the entire domain is appropriate (see Eq. (\ref{eq:norm2})); on the other hand, for the fluid we use the infinity norm because for the solution of Euler equations we use a background atmosphere density, then a norm resulting from an integral in the whole domain is affected by the atmosphere and a norm resulting from the value at a single cell (see Eq. (\ref{eq:norminfty})) is more appropriate to monitor the accuracy of the physical solution.
The time domain is such that it contains the two configurations before they abandon the domain. The SCF for both $||\rho ||_{\infty}$ and $||\Psi ||_2$ lies within $2.07$ and $2.45$, which indicates self-convergence of the numerical solutions.

The convergence properties of these three tests indicate that the ingredients involved in the program, namely, the evolution of hydrodynamical variables of the fluid, the evolution of the wave function of the BEC and the Poisson solver, work fine together, and in agreement with the accuracy of the numerical methods involved. The convergence tests of this section define the convergence regime of resolution, that we locate from the finest in our tests $\Delta x=0.25$ up to the coarsest 0.5. Simulations within this regime are said to lie within the convergence regime. In the following section we explore simulations using resolutions in most cases within the convergence regime or better, except in two cases, where we use a coarser resolution 0.625, slightly out of the convergence regime explored in tests, used to illustrate morphological properties.

\section{Preliminary applications}
\label{sec:applications}

In this section we test the code with some potential applications that still simple but near more realistic astrophysical  scenarios.

\subsection{Mergers of BEC-cores and clouds of gas }
\label{sec:mergers}

A particularly interesting scenario is that of the interaction between dark matter and gas. Assuming the BEC structures dominate the gravitational field of a potential well and a cloud of gas approaches this well, one can picture whether it is possible to track the dynamics of the fluid coupled to the dynamics of the BEC. For this we produce some simulations of the merger between a BEC core and a gas cloud modeled with a polytropic and a pressure-less gas. The idea is to show the capabilities of the code to explore these scenarios and to illustrate the diagnostics tools in the code, that help monitoring the evolution properties of the BEC+fluid system.

This basic model assumes that BEC cores, being late-time attractors, are stationary ground state solutions of the SP system \citep{GuzmanUrena2006,Schive:2014dra,Mocz:2017wlg}, that form and produce gravitational potential wells, where later on baryonic clouds fall into.

\subsubsection{Initial conditions}

The BEC-core is assumed to be the ground state equilibrium configuration used in the first test above. On the other hand, since there is no prescription on the properties of the baryonic gas, we arbitrarily use two fluid configurations, first  a TOV star and second a ball of dust.

Likewise in the examples above, the BEC configuration has a central density $\frac{1}{4\pi}$ which sets its mass to $M_{BEC}=2.06$ in code units \citep{GuzmanUrena2004}. The mass of the fluid configuration $M_{gas}$ is assumed to be either 10\% and 50\% of the mass of the BEC configuration and can be a TOV star or a dust-ball, and we define the mass ratio $\lambda=M_{gas}/M_{BEC}$.  

The initial position and momentum of the two configurations are such that the center of mass lies at the coordinate origin and the initial positions of the centers of the BEC-core and gas configurations are $(\lambda x_0,\lambda y_0,0)$ and $(x_0,y_0,0)$ respectively.

\begin{figure}
\includegraphics[width=8cm]{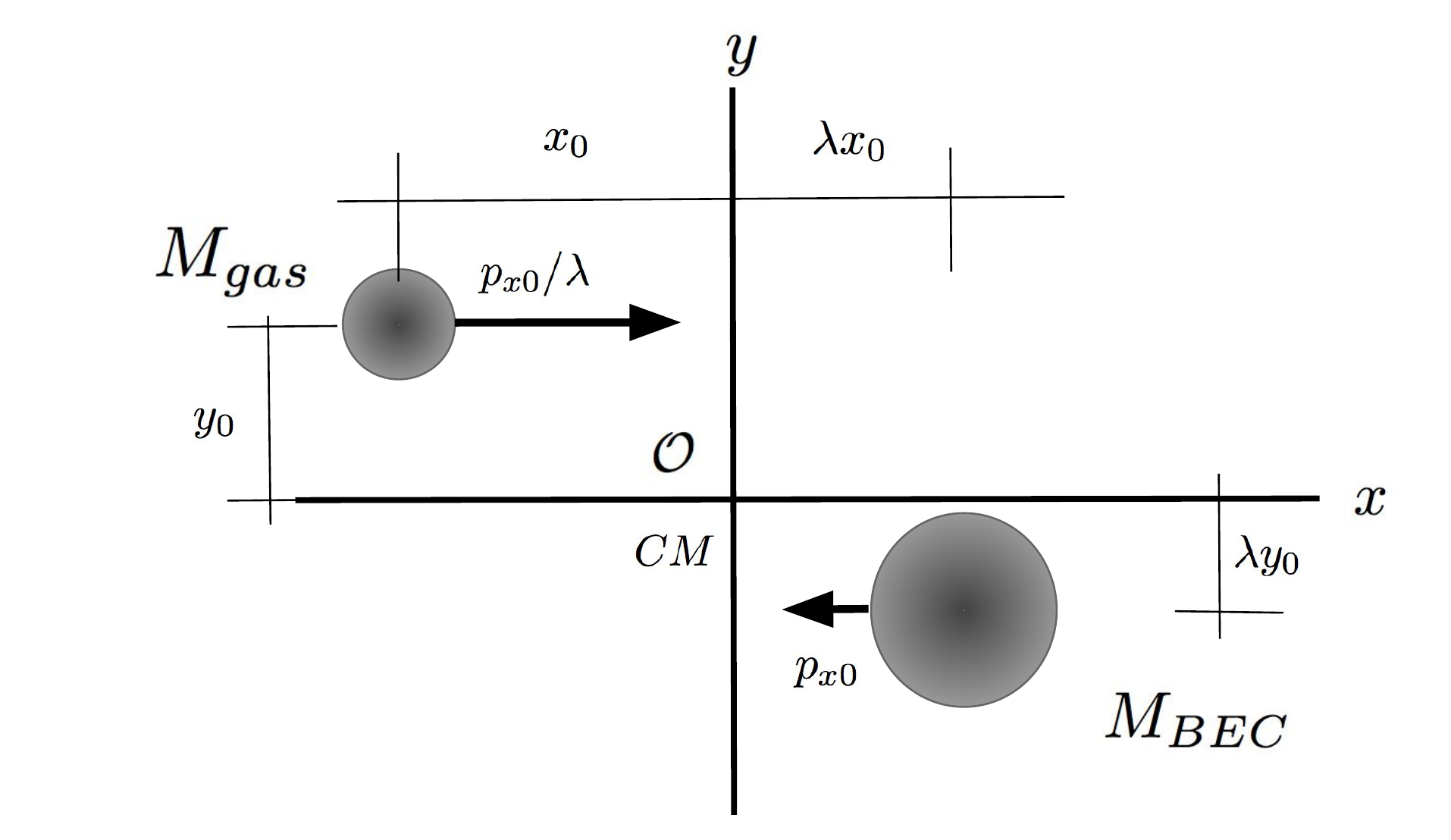} 
\caption{\label{fig:ics} Scheme illustrating the initial conditions of the BEC and fluid merger on the $xy-$plane, with the center of mass at the coordinate origin.}
\end{figure}

Initial momentum is applied along the $x-$direction in such a way that the center of mass remains nearly at the center of the domain. If the initial momentum of the BEC core is set to $\vec{p}_{0,BEC}=(p_{0x} / M_{BEC},0,0)$, then the initial momentum of the gas is $\vec{p}_{0,gas}=-\vec{p}_{0,BEC}$. The momentum of the BEC core is implemented by rescaling the wave function of the equilibrium configuration $\Psi_{eq}\rightarrow e^{-i p_{0x}x / M_{BEC}}\Psi_{eq}$.  The reason is that the wave function resulting from the product of the equilibrium wave function and a plane wave along $x$, is such that the expected value of the linear momentum operator $\hat{p}_x$ is precisely $\expval{ \hat{p}_x } = i\int \Psi^* \frac{d\Psi}{dx}d^3x = p_{0x}$. In this particular merger problem we set the momentum along $y$ and $z$ directions to $\expval{\hat{p}_y} = \expval{\hat{p}_z} = 0$. On the part of the fluid, since the spherical TOV or dust distribution has initial zero velocity with respect to its center, the velocity of the gas is applied to each volume element of the star $\vec{v}_{0,gas}=(-\frac{p_{0x}}{M_{gas}},0,0)$. A picture that illustrate these initial conditions appears in Figure \ref{fig:ics}.

Summarizing, initial positions and velocities of the system are parametrized by the impact parameter $y_0$ of the gas ball and the initial horizontal momentum $p_{0x}$ of the BEC core. These initial conditions are used to integrate the SEP equations (\ref{eq:EGPPrho})-(\ref{eq:EGPPPoisson}) as an Initial Value Problem.

\subsubsection{Parameter space}

We now present a sample of BEC+fluid mergers. We show results for two main scenarios: i) an ideal gas with adiabatic index $\gamma=5/3$ and ii) a dust-ball with $p=0$. For each of these scenarios we designed a number of simulations with different values of impact parameter $y_0$ and initial momentum $p_{0x}$ described above. We also assume two values of the mass ratio $\lambda =0.1$ and $0.5$.

\subsubsection{Ideal gas}

We explore the merger of a BEC core with a TOV star with  adiabatic index $\gamma = 5/3$.
The parameter space covers the mass ratio $\lambda = 0.1, 0.5$,  impact parameter $y_0= 10,15,20$ and the initial TOV speed $v_{x0} = 0.1$ and $0.2$. We simulate the evolution of these encounters in the numerical domain $[-100,100]\times[-100,100]\times[-50,50]$ using FMR with two refinement levels with $\Delta x = \Delta y = \Delta z = 10 / 16$ in the second refinement and $C=0.1$ is used at the highest resolution refinement level.

\begin{figure*}
\centering
\subfigure[Virialization]{\includegraphics[width=7.5cm]{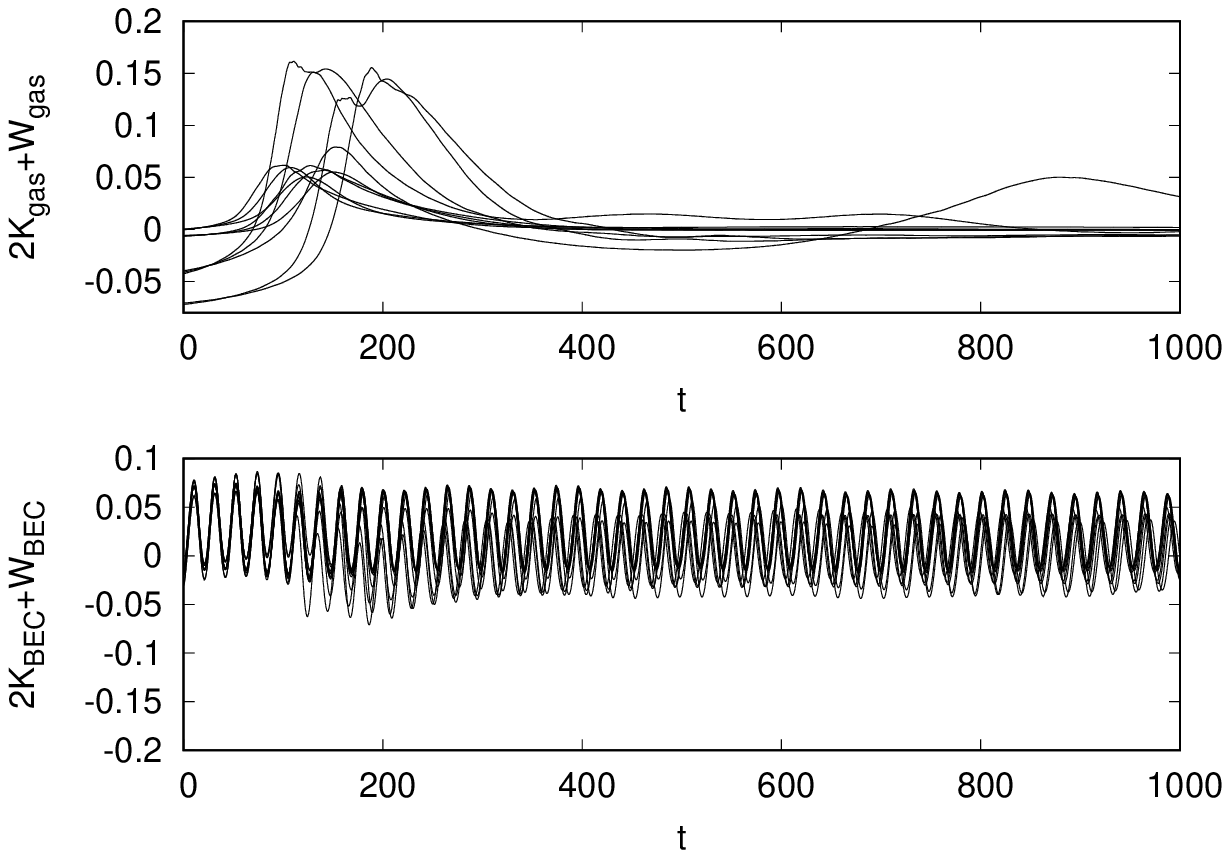}}\hspace{1mm}
\subfigure[Total energy]{\includegraphics[width=7.5cm]{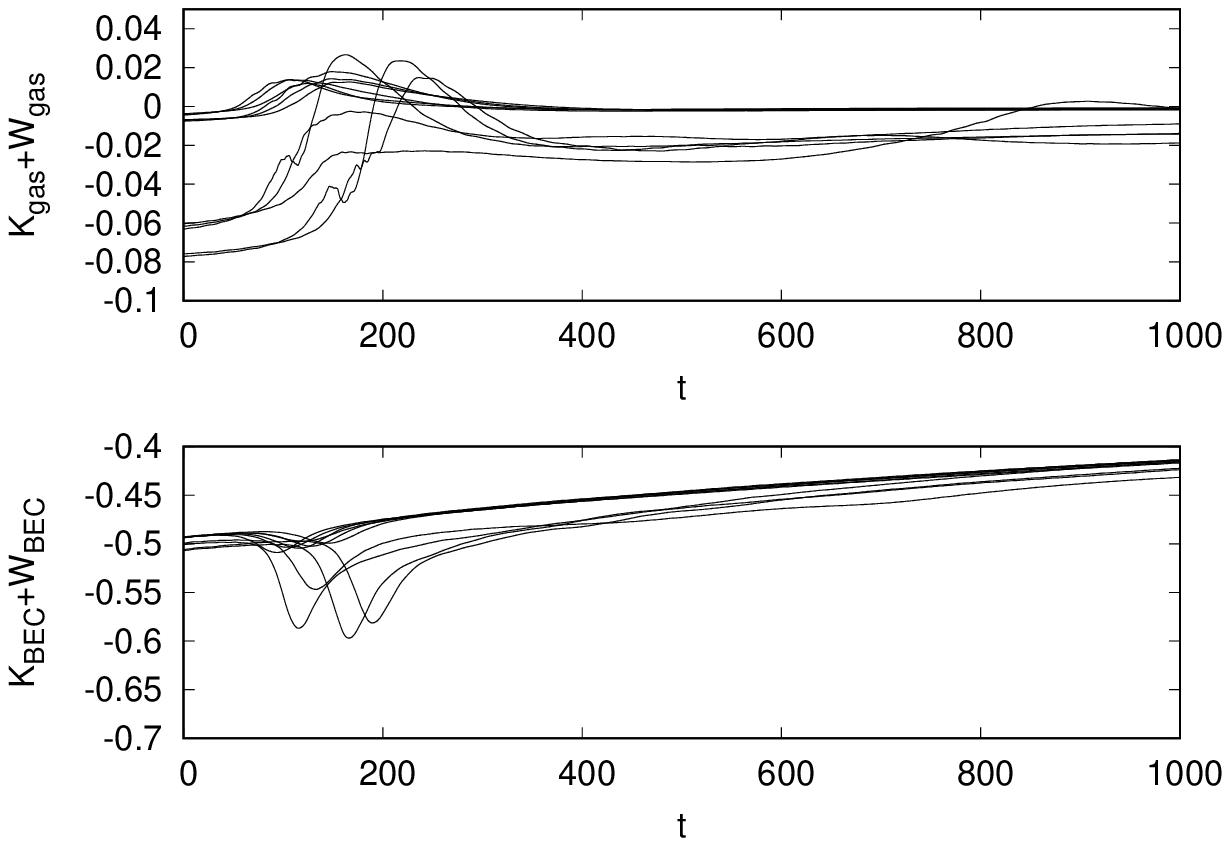}}\vspace{1mm}
\subfigure[Angular momentum]{\includegraphics[width=7.5cm]{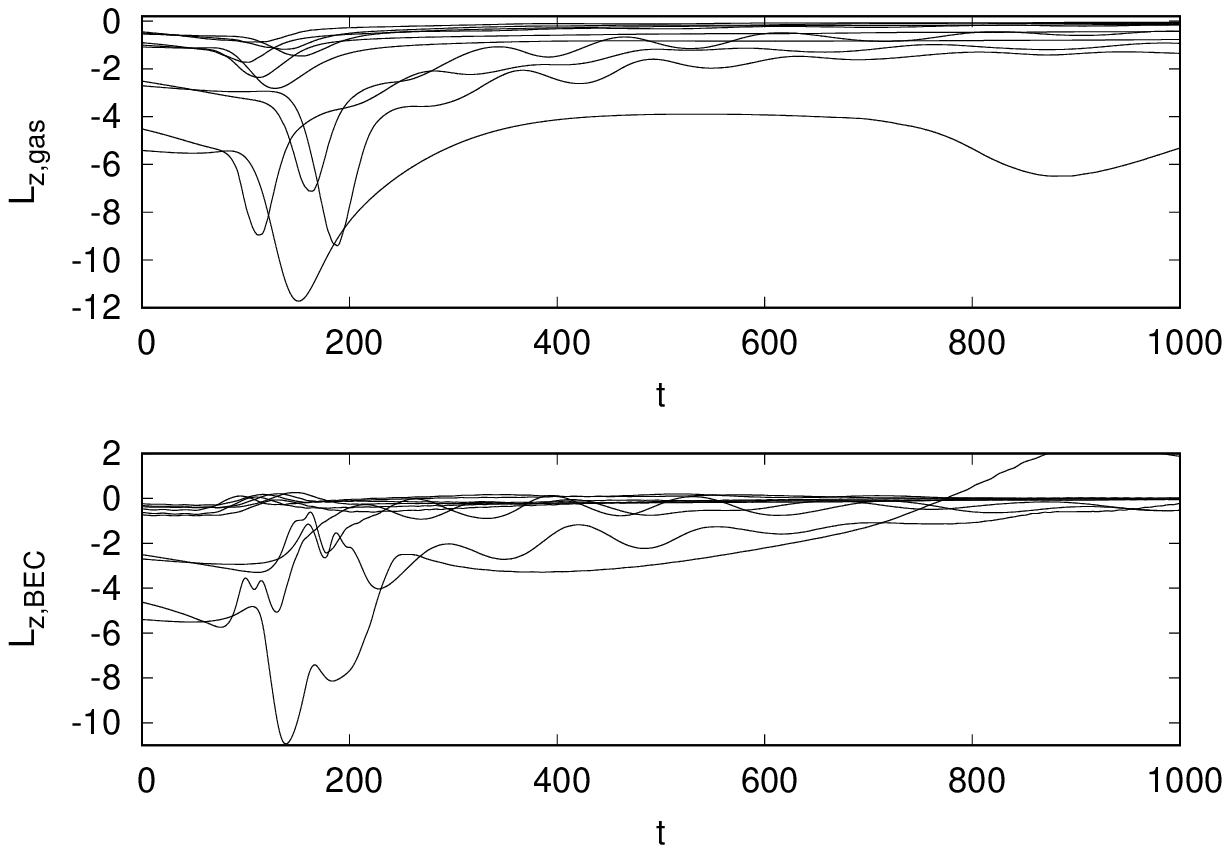}}\hspace{1mm}
\subfigure[Mass ratio]{\includegraphics[width=7.5cm]{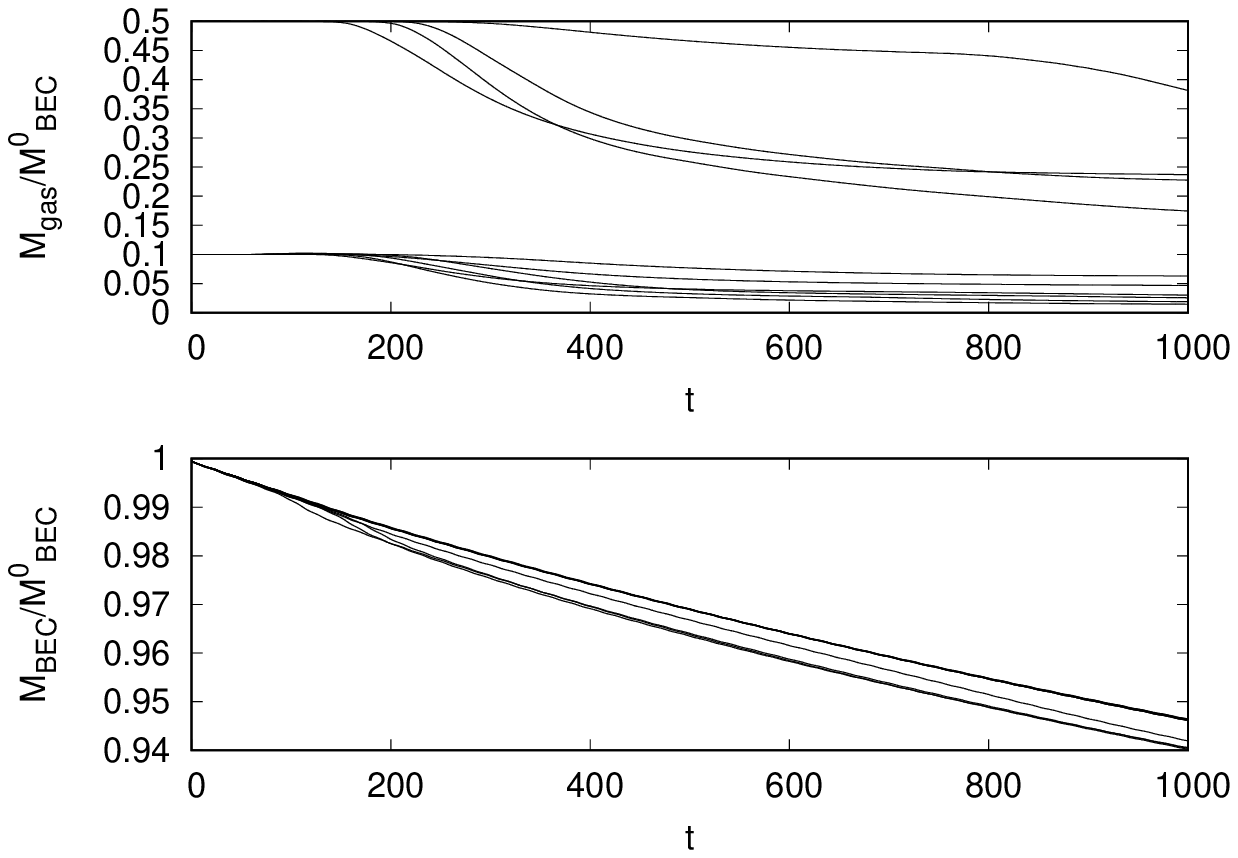}}\vspace{1mm}
\caption{Merger  of the BEC-core and the TOV star. For all the combinations of parameters we show in (a) the quantity $2K+W$ for both the BEC and the gas,  in $(b)$ the total energy $K+W$, in (c) the total angular momentum in the direction $z$,  and finally  in (d) the mass normalized with the initial mass of the BEC, $M^0_{BEC}\approx 2.06$ as a function of time.}
\label{fig: conservatives}
\end{figure*}

The dynamics of the system is such that the gas tends to orbit around the core clockwise. Once the TOV star encounters the  core, the pressure of the gas fragments the star and the fluid circulates around the BEC core from the two sides, which triggers the formation of shocks. Diagnostics of configurations and dynamical behavior of each component include the virialization, total energy, release of angular momentum and mass. In  Figure \ref{fig: conservatives} we show these diagnostic variables defined as in equations (\ref{eq:BECM})-(\ref{eq:gasLz}) until $t\sim 1000$, for all the combinations of parameters. 

General observations are the following.
The gas relaxes towards a virialized state since $2K_{gas}+W_{gas}$ tends to zero in all cases and its mass  also stabilizes. This indicates that the fluid relaxes and settles down.
The BEC oscillates around a virialized state with $2K_{BEC}+W_{BEC}$  around zero, whereas the mass $M_{BEC}$ decreases in time since the sponge absorbs BEC density approaching the boundary.
The angular momentum of the gas stabilizes around a finite value whereas the angular momentum of the BEC does as well in most cases.
The energy of the gas $E_{gas}=K_{gas}+W_{gas}$ also settles, whereas that of the BEC $E_{BEC}=K_{BEC}+W_{BEC}$ in this time scale still does not approximate a constant value.

\begin{figure}
\centering
\includegraphics[width=8cm]{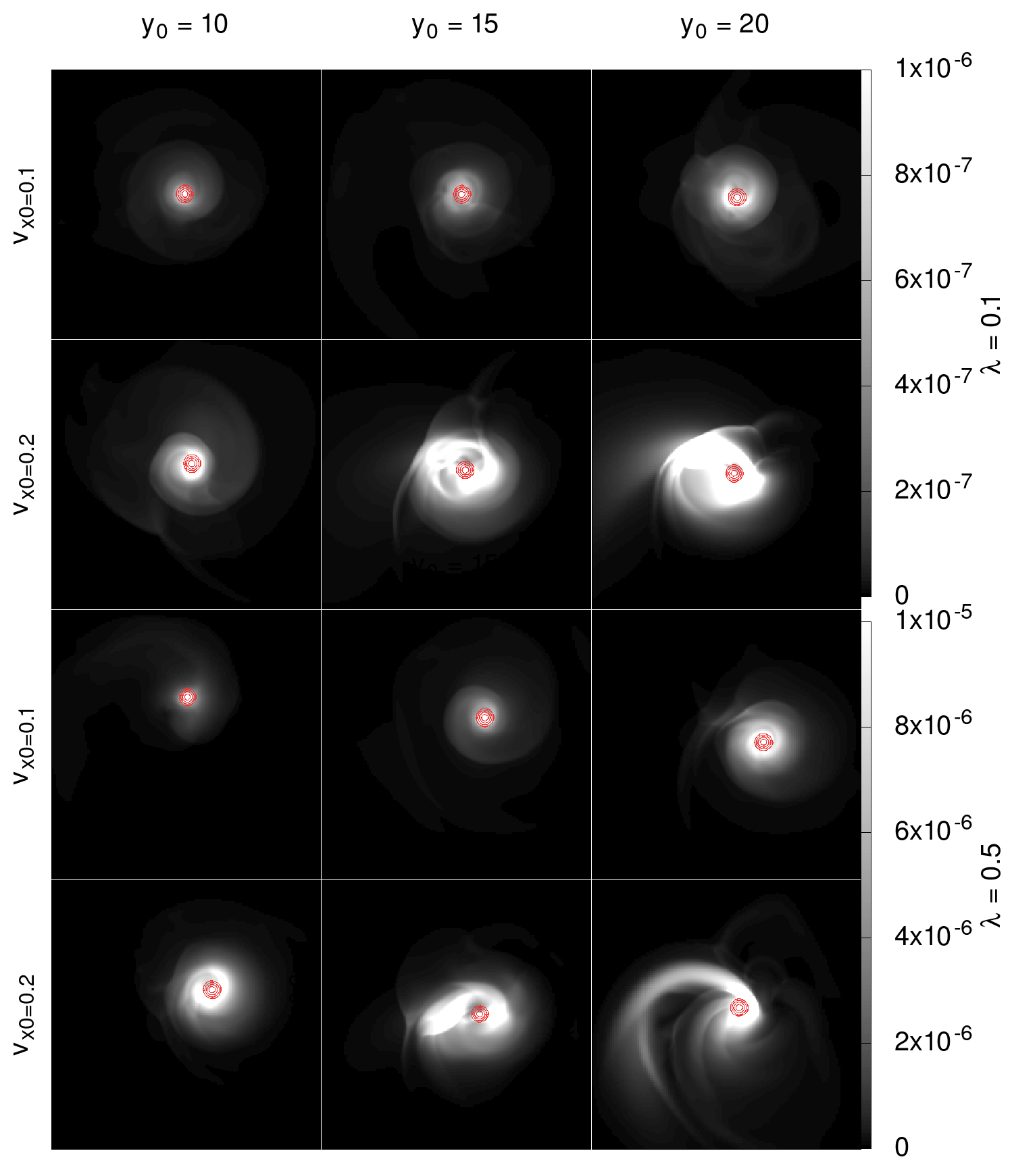}
\caption{Merger  of the BEC-core and the TOV star. Snapshot by $t\sim1000$ of the gas density $\rho$ and contours of $|\Psi|^2$ at the plane $z=0$ for all combinations of parameters.}
\label{fig: all data gas}
\end{figure}

Figure \ref{fig: all data gas} shows a late-time snapshot of the BEC and gas densities for all combinations of parameters.  These simulations show a gas density distributed on a disk on the equatorial plane and various morphologies, in some cases with spiral arms and in some others more uniform disks. The motion of the gas is permanent and we only capture a snapshot. 

\begin{figure}
\includegraphics[width=3.5cm]{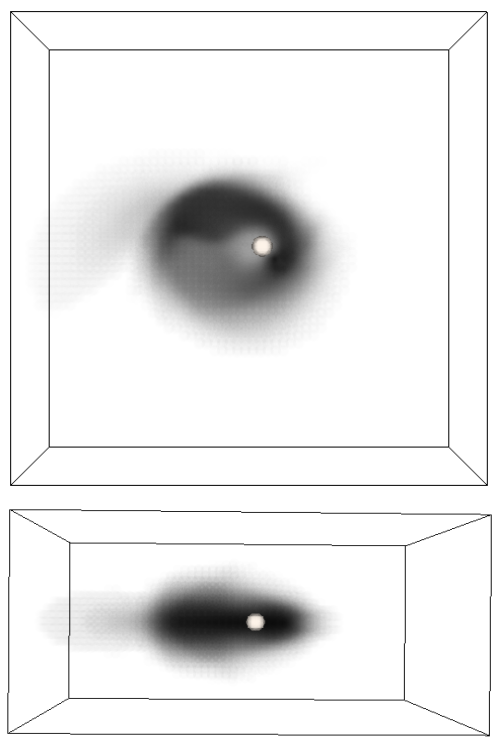}
\includegraphics[width=3.5cm]{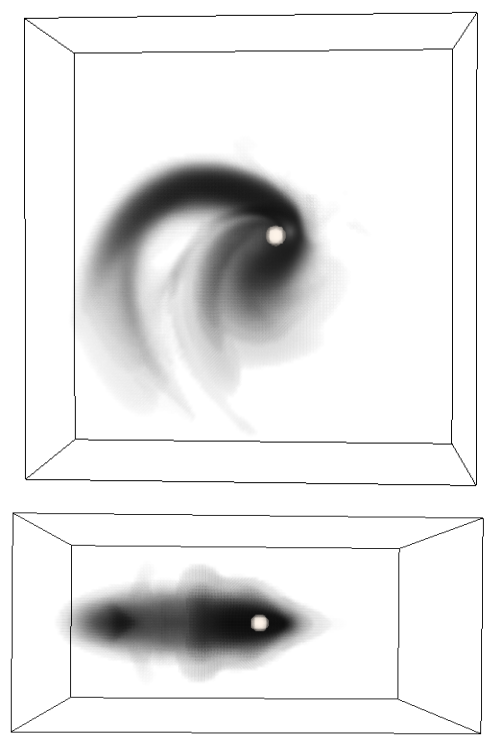}\vspace{1mm}
\caption{Late-time snapshot of the merger between the BEC-core and the TOV star for the case $v_{x0}=0.2$ and $y_0=20$. (Top) Density of the ideal gas as seen from the $z-$axis and (bottom) the gas density as viewed from the $x-$axis. On the left we show the case of mass ratio $\lambda=0.1$ and on the right $\lambda=0.5$.}
\label{fig: 3d gas}
\end{figure}

These simulations can be viewed more closely. For example, in Figure \ref{fig: 3d gas} we show the density of the BEC and the gas, from top and side perspectives, for the particular case of $y_0 = 20$, $v_{x0}=0.2$ and mass ratio $\lambda = 0.1,0.5$. The gas distributes with high density near the equatorial plane and the development of an overdensity similar to a bulge also appears. The BEC core on the other hand  distributes nearly spherically.

\subsubsection{Dust}

We now explore the merger between a BEC core and a dust-ball using the same parameter space as in the previous scenario. Unlike for the TOV case, which is in itself an equilibrium configuration, for dust we simply use the TOV density profile and set the pressure to zero all the way during the evolution. Non-trivial dynamics happens prior to merger, namely the free-fall collapse of the dust into its own gravitational potential.

In this pressure-less case we use the numerical domain $D = [-50,50]\times[-50,50]\times[-25,25]$ smaller than in the previous case and solve using unigrid mode with resolution $\Delta x=\Delta y=\Delta z=10/16$. For comparison, the parameter space includes the same orbital parameters used for the TOV star, that is, with $\lambda = 0.1, 0.5$,  impact parameter $y_0= 10,15,20$ and initial dust-ball speed $v_{x0} = 0.1$ and $0.2$.

In the diagnostics of this scenario, similar results to the previous ideal gas case were found for virialization, total energy, mass and angular momentum. 
Again, the fluid virializes, acquires a finite angular momentum and its mass stabilizes, whereas the BEC radiates the angular momentum in most cases, does not virialize but oscillates around a virial state and loses mass continuously due to ejection during the merging process.

\begin{figure}
\includegraphics[width=8cm]{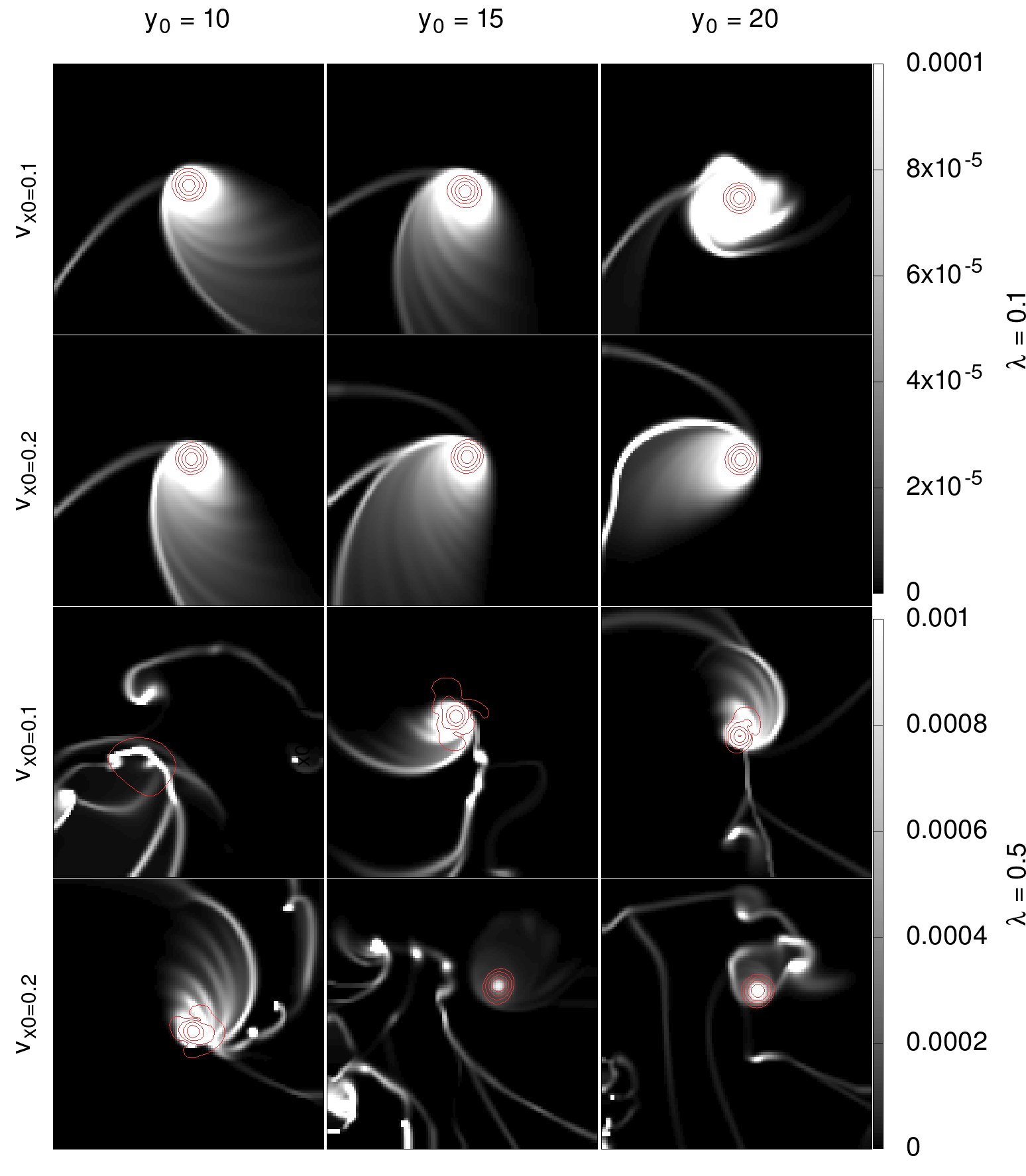}
\caption{Late-time snapshot, by time $t \sim 1000$, of the merger between the BEC-core and the dust-ball. 
Gas density $\rho$ in white and contours of $|\Psi|^2$ on the plane $z=0$ at for all combinations of parameters.} 
\label{fig: all data dust}
\end{figure}

Figure \ref{fig: all data dust} shows a late-time snapshot of the merger between the densities of the BEC core and the dust-ball for all combinations of parameters. In this case, again the fluid density concentrates in the equatorial plane, and even if there is no pressure, high density spiral arms are formed in some cases. 
The case with $\lambda=0.5,~v_{x0}=0.1,~y_0=10$, illustrates how the quick collapse of the dust before disruption produces a deep gravitational potential that distorts also the BEC and eventually both components are expelled from the domain. Unlike the ideal gas case, in various of the configurations for dust in Figure \ref{fig: all data dust}, the fluid density distributes in fringed structures. 

\begin{figure}
\includegraphics[width=3.5cm]{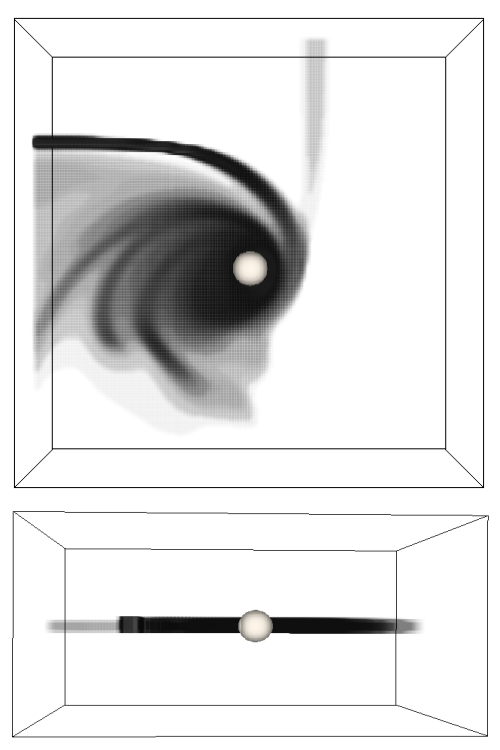}
\includegraphics[width=3.5cm]{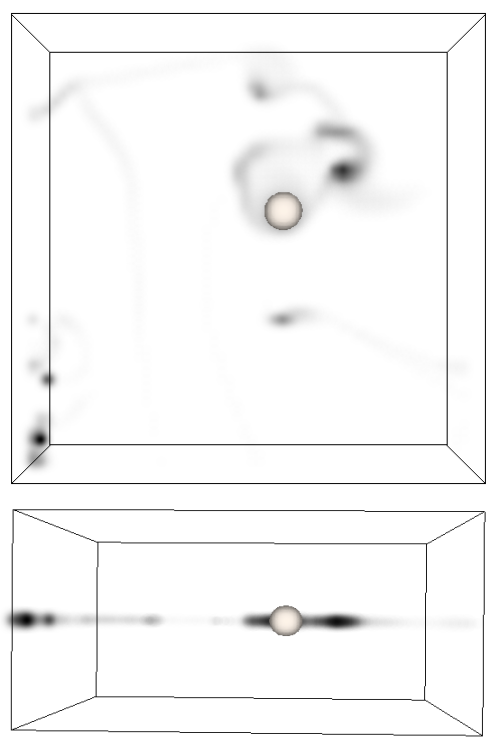}
\caption{Late-time snapshot of densities in the merger between the BEC-core and the dust-ball. (Top) As seen from the $z-$axis, and (bottom) as viewed from $x-$axis, for the case $v_{x0}=0.2$ and $y_0=20$. On the left we show the case of mass ratio $\lambda=0.1$ and on the right $\lambda=0.5$.}
\label{fig: 3d dust}
\end{figure}

A late-time view of densities for the case $y_0 = 20$, $v_{x0}=0.2$ with mass ratios $\lambda = 0.1,0.5$ is shown in Figure \ref{fig: 3d dust}. Likewise in the ideal gas case, the dust distributes on spiral arms, this time on a thin distribution concentrated near the equatorial plane due to the lack of pressure. The BEC does again become nearly spherical. Notice that the ideal gas in Figure \ref{fig: 3d gas} for the same initial dynamical parameters, distributes in a thicker distribution near the equatorial plane compared with the thin distribution of the dust, an expected behavior due to  pressure effects.

\subsection{Equilibrium BEC+fluid configurations}

In this subsection we present the construction of equilibrium configurations made of BEC and fluid. Later on we explore the capabilities of the code in scenarios involving mergers between two of these equilibrium configurations. The examples also illustrate how the code could handle scenarios involving structures made of dark and baryonic matter.

\subsubsection{Equilibrium configurations}
\label{subsub:equlibrium}

We assume that the fluid is in hydrostatic equilibrium, that the wave function is stationary, and that the system is spherically symmetric. Under these considerations the SEP system is reduced to the following set of ordinary differential  equations

\begin{equation}
\dfrac{dM}{dr} = 4\pi r^2 \left(\rho + \psi^2\right), 
\label{eq: mass stationary}
\end{equation}

\begin{equation}
\dfrac{dp}{dr} = -\rho \dfrac{M}{r^2},
\label{eq: pressure stationary}
\end{equation}

\begin{equation}
\dfrac{d^2\psi}{dr^2} + \dfrac{2}{r}\dfrac{d\psi}{dr} = 2\left(V+\omega\right)\psi,
\label{eq: Schrödinger stationary}
\end{equation}

\begin{equation}
\dfrac{dV}{dr} = \dfrac{M}{r^2},
\label{eq: potential stationary}
\end{equation}

\noindent where $\Psi =  \psi e^{-i \omega t}$, $\psi=\psi(r)$ is a real, node-less, positive function, $\omega$ is a constant of integration and $M(r)=M_{BEC}(r)+M_{gas}(r)$ is the total mass of the system resulting from the integration of the densities $\rho$ and $\psi^2$ up to a radius $r$. The system is closed with a polytropic equation of state. A solution of the system can be uniquely determined under the boundary conditions $M(0)=0$, $p(0) = K\rho_c^\gamma$, $\psi(0) = \psi_c$, $\psi(r)=\psi'(r)\to 0$ when $r\to\infty$ and $V(0)=V_c$, where a solution is determined by the value of the central density $\rho_c$ and the central value of the wave function $\psi_c$, once the value of the polytropic constant $K$ and the polytropic index $n=1/(\gamma-1)$ are given. The value of $V_c$ can be chosen arbitrarily, since the potential can be rescaled once the total mass of the system with a monopolar boundary condition is imposed. In this way the system (\ref{eq: mass stationary}-\ref{eq: potential stationary}) is translated into an eigenvalue problem for the eigenvalue $\omega$, which we solve numerically using the shooting method \citep{nr}. 
The parameter that characterizes the amount of BEC and fluid in these solutions is the mass ratio $\lambda$ defined as

\begin{equation}
\lambda = \dfrac{M_{gas}}{M_{BEC}} = \dfrac{\int_{0}^\infty \rho r^2 dr}{\int_{0}^\infty |\psi|^2 r^2 dr}.
\label{eq: lambda}
\end{equation}

\begin{figure}
\centering
\includegraphics[width=7.5cm]{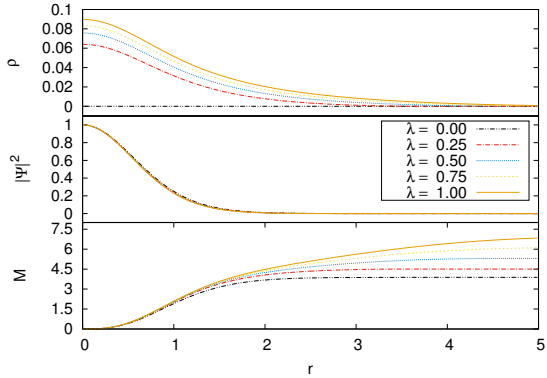}
\caption{Equilibrium solutions of the SEP system for values of the mass ratio $\lambda = M_{gas} / M_{BEC} = 0.0$, 0.25, 0.5, 0.75 and 1.0 for a polytrope with polytropic index $n=1.5$ and polytropic constant $K=10$.}
\label{fig: equilibrium solutions}
\end{figure}

\noindent We show examples of solutions for various values of $\lambda$ in Figure \ref{fig: equilibrium solutions}, where we use $K=10$, $n=1.5$, $\psi_c=1$. The central density $\rho_c$ is found in such a way that the factor $\lambda$ takes the values $0$, $0.25$, $0.50$, $0.75$ and $1$ where $\lambda = 0$ corresponds to the case when there is no fluid, that is $\rho_0 = 0$ and the ground state equilibrium solution of the Schr\"oedinger-Poisson system in \citep{GuzmanUrena2004} is recovered.

\subsubsection{Collision between two of these configurations}

We collide two of these equilibrium solutions parameterized by the mass ratios $\lambda_1$ and $\lambda_2$, with total masses $M_1$ and $M_2$ respectively, placed in positions $(x_0,y_0,0)$ and $-M_1 / M_2 (x_0 ,y_0,0)$. We study  mergers between these configurations by injecting an initial velocity $v_{x0}$ to the solution with mass $M_1$ and an initial velocity $- M_1 v_{x0} / M_2$ to the solution with mass $M_2$.

Using the equilibrium configurations with $\lambda_1 = 0.25$ and $\lambda_2 = 0.25, 0.50$ placed at $x_0=-5$, we study  two scenarios i) head-on collisions with velocities $v_{x0}=1, 2,3,4,5,6$ and ii) fusions with impact parameter $y_0=2.5,5$,  and initial velocity $v_{x0}=0.1$.

\subsubsection{Head-on collision}

We evolve the system in the cubic domain $D = [-20,20]^3$ using two refinement levels with the finest resolution $\Delta x = 1/6$ and a courant number $C = \Delta t /\Delta x^2 = 0.0225$ in the fine mesh. \\

\begin{figure}
\centering
\includegraphics[width=8.5cm]{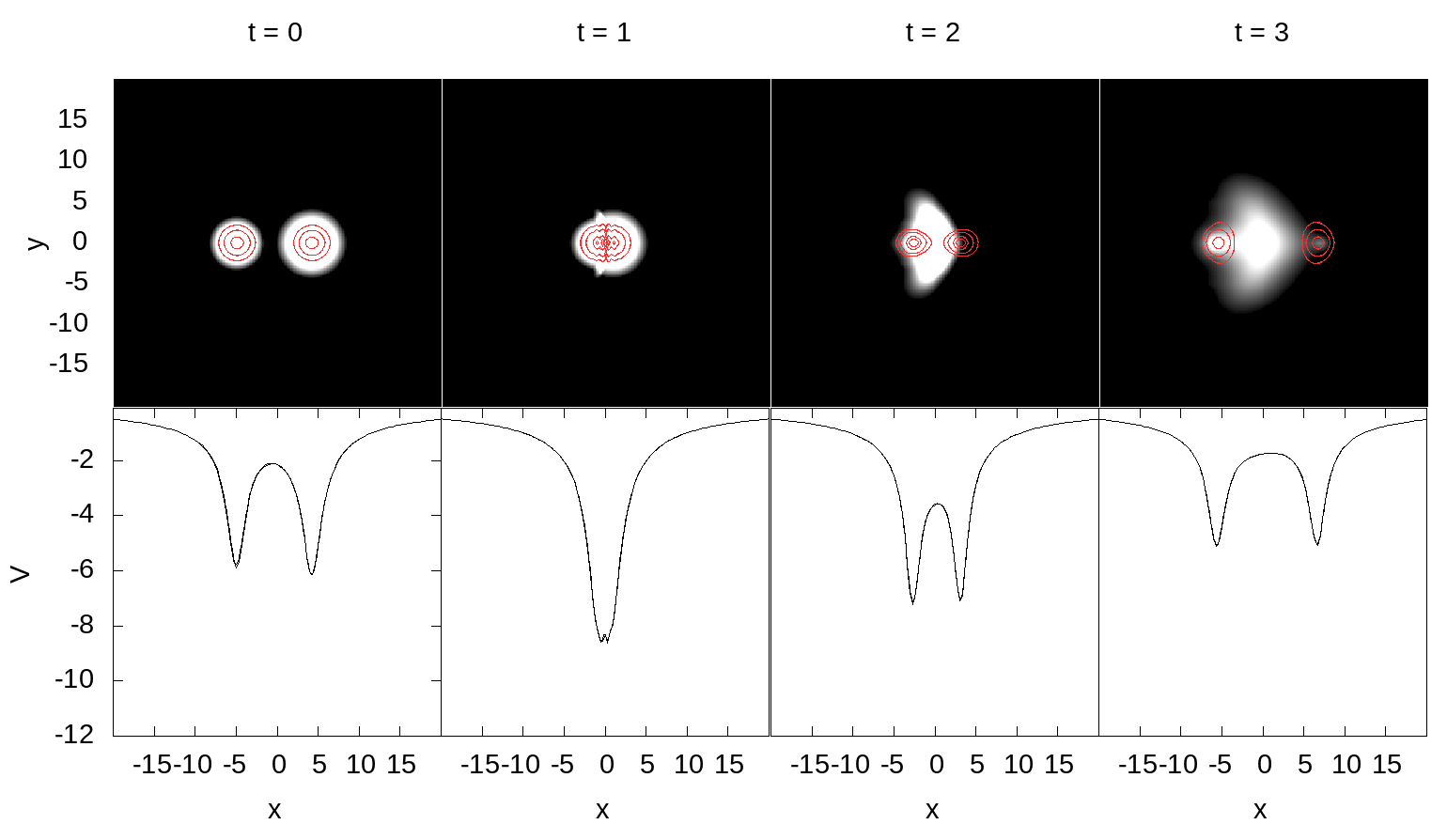}
\caption{(Top) snapshots of the head-on merger for initial parameters $v_{x0} = 4$ and $\lambda_1 = \lambda_2/2 = 0.25$, where the BEC density is represented with isocontours in red and the gas density with a grayscale map in the $z=0$ plane. (Bottom) the gravitational potential along the $x$-axis.}
\label{fig: evolve headon}
\end{figure}

The evolution of one of these configurations with parameters $\lambda_{2} = 0.5$ and $v_{x0} = 4$ is shown in  Figure \ref{fig: evolve headon}. In the upper part we show various snapshots of the isocontours of the BEC density profile, and a scaled gray map for the gas density profile on the $z=0$ plane. At the bottom we show the evolution of the gravitational potential along the $x$-axis, in which it can be seen how the configuration with $\lambda_2 = 0.5$ produces a deeper potential well than the configuration with $\lambda_1 = 0.25$ at initial time. By time $t=1$ the interference pattern produced by the interaction between the BEC densities begins to reflect in the gravitational potential, and will finally have effects on the gas distribution. At a later time the BEC and gas densities have already collided and are beginning to move away from each other, leaving part of the gas traveling with the BEC and part behind the BEC. 

\begin{figure}
\centering
\includegraphics[width=8cm]{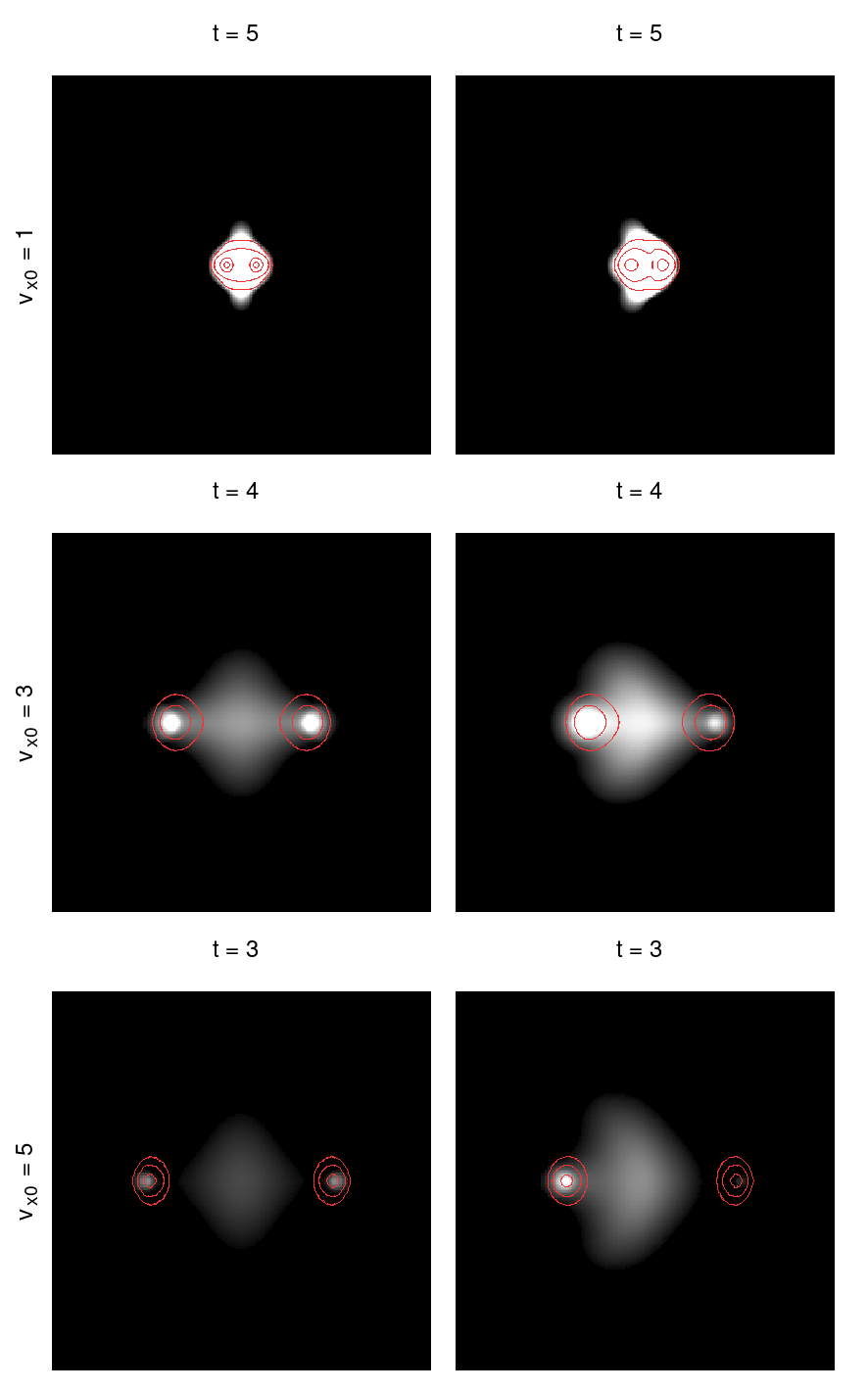}
\caption{Gas density in grayscale map and BEC density isocontours on the plane $z=0$, for various combinations of parameters. In each case we draw the snapshot corresponding to a particular time, at the time when the two configurations either merge or pass through each other. Notice that in some cases, the gas acquires a bullet head shape which was explored for  models of Bullet Cluster type of systems assuming Cold Dark Matter in \citep{BulletCDM}.}
\label{fig: density headon}
\end{figure}

In Figure \ref{fig: density headon} we show a snapshot for each combination of initial conditions. It can be seen that as the velocity increases, the BEC isocontours tend to separate from the gas density, which is possibly delayed due to the effects of the pressure of the gas itself. The later cases show the offset between BEC dark matter and the cloud of gas, so as the bullet shaped core and head of the gas, that have been investigated at the Bullet-Cluster scale within the Cold Dark Matter model \citep{BulletCDM}.

\subsubsection{Mergers with orbital momentum}

Now we solve the system in the cubic domain $D=[-20,20]^3$ using a single refinement level, with resolution $\Delta x = 1/3$ and courant number $C=0.1$.

\begin{figure}
\centering
\includegraphics[width=8.5cm]{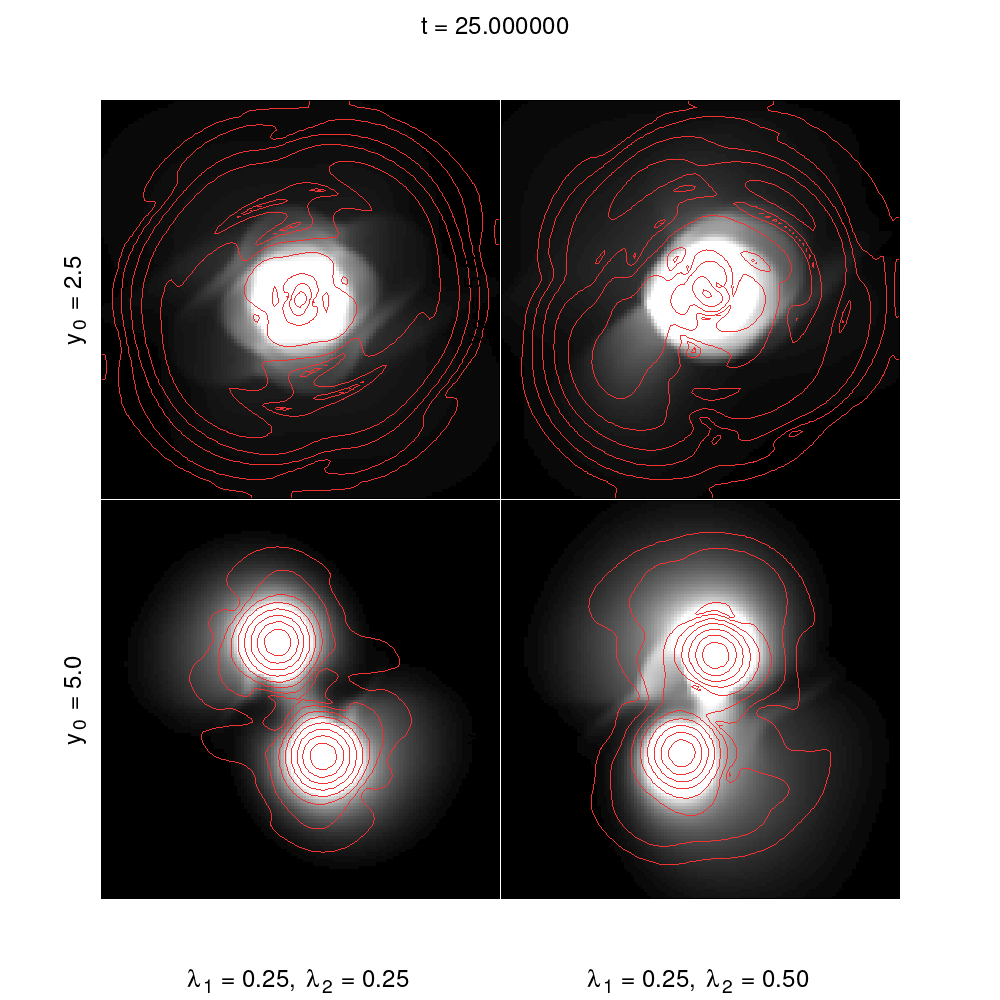}
\caption{Snapshot of mergers by $t=25$. Gas (grayscale map) and BEC (red isocontours) densities in the plane $z=0$ at time $t=25$ for the mergers of the BEC-gas configurations for all the parameters explored. 
The four cases correspond to parameters $y_0 = 0.25, 0.5$, $\lambda_{2}=0.25, 0.5$ and $v_{x0} = 0.1$. }
\label{fig: density}
\end{figure}

\begin{figure}
\centering
\includegraphics[width=8cm]{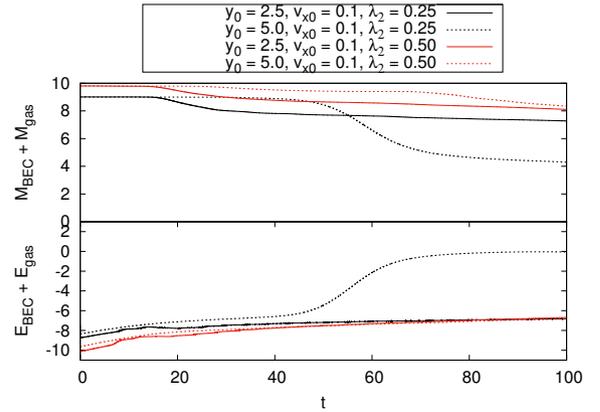}
\caption{Total mass and energy as a function of time for the merger between BEC-gas configurations for various combinations of mass ratio and impact parameter. Part of the total mass is lost and the total energy tends to a negative constant value, which indicates that the system is bounded. The loss of mass is due to the boundary conditions, the sponge for the wave function and outflow conditions for the fluid.}
\label{fig: diagnostic}
\end{figure}

In Figure \ref{fig: density} the density profiles of the gas and the BEC are shown at time $t=25$ on the plane $z=0$. Observe that the configurations with $y_0=2.5$, $v_{x0}=0.1$, $\lambda_2 = 0.25$ and $0.5$ merge and form a new cuasi-stationary distribution that evolves with nearly constant global mass and energy. Similar results are found for the initial conditions with $y_0=5.0$, $v_{x0}=0.1$, $\lambda_2 = 0.25,~0.5 $.

\subsection{Oscillations of the fluid on top of a BEC core}

The BEC dark matter core has a clear spectrum of oscillations, for example oscillations that appear in the convergence test of Figure \ref{fig:BECeq} correspond to the fundamental oscillation mode, which is permanently excited by the sole truncation errors of the numerical methods used for the evolution of the system. Nevertheless, when explicit perturbations of finite mass are applied to an equilibrium configuration, the same modes are excited due to the matter distribution  around, as found in \citep{Guzman2019}, which has been explored for FDM cores along with their random walks \citep{Li_2021}. 

\begin{figure}
\centering
\includegraphics[width=8cm]{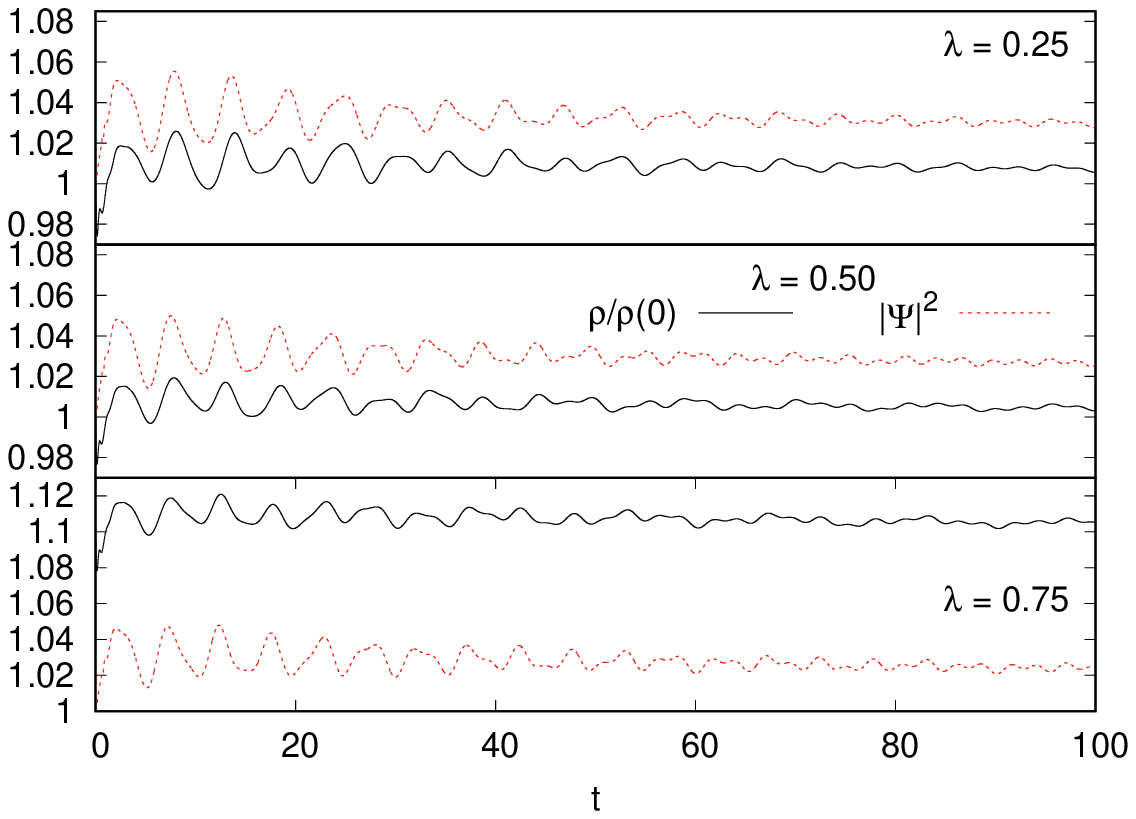}
\caption{Evolution of the central gas densities (continuous black line) and BEC (red dotted line) as function of time, of BEC+fluid equilibrium configurations for the values $\lambda = 0.25, 0.50$ and 0.75, in the upper, middle and lower panels, respectively. These figures illustrate the correlation of frequency and phase between oscillations of the BEC and the gas.}
\label{fig: evolve central density equilbrium}
\end{figure}

\begin{figure}
\centering
\includegraphics[width=8cm]{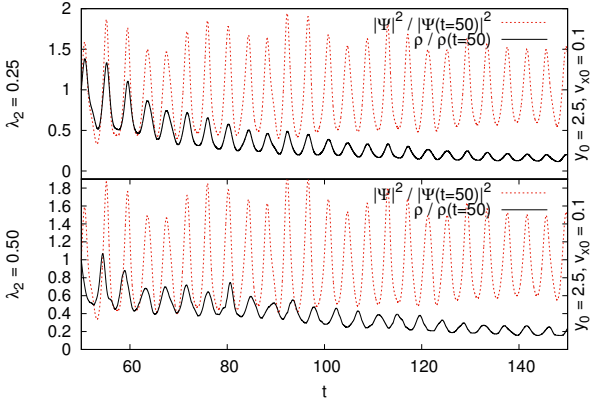}
\caption{Evolution of the central densities of the gas (solid black line) and of the BEC (red dotted line) as function of time  for the density resulting from the fusion of two equilibrium configurations. These densities are normalized with their value at $t=50$ after the merger took place. In the upper panel the case of the fusion between the initial configurations with $\lambda_2 = \lambda_1 = 0.25$, whereas in the lower panel the case with $\lambda_2 = 2\lambda_1 = 0.50 $. In the two cases the impact parameter and initial velocity are $y_0= 2.5$ and $v_{x0}=0.1$. Notice that the BEC oscillations are synchronized with those of the gas.}
\label{fig: evolve central density binary}
\end{figure}

The effects of oscillations on baryonic matter have been studied already in some scenarios, for example in order to impose restrictions on the boson mass. Cores have inherent oscillations due to the persistent interference of the surroundings \citep{GotingaBarionFDM}, and can exhibit a random walk behavior \citep{Schive_2020}. This motion can affect star clusters near galactic cores that can be overheated, disrupted \citep{Marsh_2019,GotingaBarionFDM} and scattered \citep{Hui:2016,Bar_Or_2019,Moczheating}, which illustrates the type of current analysis aimed to impose constraints to the model involving scenarios with baryonic matter. Conversely, in \citep{Gotinga2022} the feedback of baryons on the formation and density profiles of BEC dark matter cores is studied.

Motivated by these advances, we study how the code handles boson-fluid scenarios using the controlled case of the BEC+fluid  equilibrium configurations constructed in subsection \ref{subsub:equlibrium}. Figure 
\ref{fig: evolve central density equilbrium} shows the evolution of the central density of BEC+fluid  equilibrium configurations with $\lambda = 0.25, 0.50$ and $0.75$, in which we can see how the oscillations of both the BEC and the gas coincide in frequency and phase. 

The transfer of oscillations from the BEC to the gas can also be observed in the configuration resulting from the merger  between equilibrium configurations like those in Figure \ref{fig: density}. For example for the case with $\lambda_2 = 0.25$ and $\lambda_2 = 0.5$ with impact parameter $y_0 = 2.5$ and  initial speed $v_{x0} = 0.1$, the central densities of BEC and gas are shown in Figure \ref{fig: evolve central density binary}. Notice again how the oscillation of the core, which is the more massive component is transferred to the gas with potentially observable results.

\section{Discussion and conclusions}
\label{sec:conclusions}

We have described a code that solves the SEP system of equations that can help modeling the dynamics of Fuzzy Dark Matter structures interacting with a fluid, coupled through the gravitational potential of the system.

Convergence tests of the evolution with the two components separately and together indicate, that there is  second order convergence of the code that solves Schr\"odinger equation and convergence between first and second order for the fluid. Convergence of the fluid can improve in scenarios where the gas density is smooth.

As an example reminiscent of galaxy formation within the FDM scenario, we show the merger between a dark matter core and a fluid configuration. We show a sample of results for different initial conditions of mergers that give rise to configurations with different dynamical and morphological properties. In the most interesting cases we show the formation of spirals on  disk like distribution of gas together with a bulge.

We presented also the construction of BEC+fluid equilibrium configurations, that we use as to test basic scenarios of dark matter and baryon configurations. We show head-on cases with rather high velocity, that illustrate an offset between dark and baryonic matter. On the other hand, the merger of these configurations in orbit, for bounded cases, results in nearly stationary final configurations that show how FDM and baryons may coexist in single configurations with angular momentum. Finally, we also showed examples of how the oscillations of the BEC distribution and that of the gas synchronize, finding a pattern that may eventually lead to observational predictions or constraints of the fuzzy dark matter model. 

We expect that the type of simulations doable with this code can be applied and contrasted within astrophysical scenarios of formation and dynamics of the various types of galaxies, within the ultralight dark matter model phenomenology.

Further versions of the code should include accurate models of baryonic matter useable in hydrodynamical simulations that so far are implemented within structure formation simulation scenarios (e.g. \citep{mocz19,mocz19b,GotingaBarionFDM}), that use non-equilibrium models of cooling for star formation scenarios at local scales, including the radiative processes developed in \citep{CAFER}. Also cosmologically motivated FDM density profiles beyond cores should be used, for example those in \citep{ShapiroCreTail,YavetzLiHui2022}.


\section*{Acknowledgements}

This research is supported by grants CIC-UMSNH-4.9 and CONACyT Ciencias de Frontera Grant No. Sinergias/304001. 
Iv\'an Alvarez receives support from the CONACyT graduate scholarship program under the CVU 967478.
The runs were carried out in the Big Mamma cluster of the Laboratorio de Inteligencia Artificial y Superc\'omputo, IFM-UMSNH.


\section*{Data availability}

The data and code underlying this article will be shared on reasonable request to the corresponding author.\\


\bibliographystyle{mnras}

\bibliography{BECDM} 



\bsp	
\label{lastpage}
\end{document}